\documentclass{cta-author}

\usepackage{tikz}
\usepackage{tikz-3dplot}
\usepackage{amssymb}
\usepackage{algorithm}
\usepackage{algpseudocode,bm,multirow}
\usepackage{mathrsfs} 
\usepackage{balance}
\usepackage{xcolor}
\usepackage{colortbl}
\usepackage{lipsum}
\usepackage{mathtools, cuted}
\usepackage{lipsum, color}
\usepackage{url}
\usepackage{multirow}
\usetikzlibrary{shapes.geometric,arrows}
\usetikzlibrary{decorations.pathreplacing}

\begin{document}

\supertitle{}

\title{On the Conditioning of the Spherical Harmonic Matrix for Spatial Audio Applications}

\author{\au{Sandeep Reddy C$^{1\corr}$}, \au{Rajesh M Hegde$^{2}$}}

\address{\add{1}{Department of Electrical Engineering, Indian
Institute of Technology, Kanpur 208016, India}
\add{2}{Department of Electrical Engineering, Indian Institute of Technology, Kanpur 208016, India}
\email{csreddyiitk@gmail.com, rhegde@iitk.ac.in}}

\begin{abstract}
In this paper, we attempt to study the conditioning of the Spherical Harmonic Matrix (SHM), which is widely used in the discrete, limited order orthogonal representation of sound fields.  SHM's has been widely used in the audio applications like spatial sound reproduction using loudspeakers, orthogonal representation of Head Related Transfer Functions (HRTFs) etc. The conditioning behaviour of the SHM depends on the sampling positions chosen in the 3D space. Identification of the optimal sampling points in the continuous 3D space that results in a well-conditioned SHM for any number of sampling points is a highly challenging task. In this work, an attempt has been made to solve a discrete version of the above problem using optimization based techniques. The discrete problem is, to identify the optimal sampling points from a discrete set of densely sampled positions of the 3D space, that minimizes the condition number of SHM. This method has been subsequently utilized for identifying the geometry of loudspeakers in the spatial sound reproduction, and in the selection of spatial sampling configurations for HRTF measurement. The application specific requirements have been formulated as additional constraints of the optimization problem. Recently developed mixed-integer optimization solvers have been used in solving the formulated problem. The performance of the obtained sampling position in each application is compared with the existing configurations. Objective measures like condition number, D-measure, and spectral distortion are used to study the performance of the sampling configurations resulting from the proposed and the existing methods. It is observed that the proposed solution is able to find the sampling points that results in a better conditioned SHM and also maintains all the application specific requirements.
\end{abstract}

\maketitle

\section{Introduction}\label{sec1}

Signals measured in real environments that are spatially dependent usually undergo perturbation errors due to the inaccurate placement of measurement setups.  The modern signal processing techniques extract useful information from the measured signals by transforming in to various domains like Fourier, Wavelets, Spherical Harmonics etc \cite{oppenheim2010discrete,rafaely}. Generally, this transforms involve signal representations in terms of continuous orthogonal basis functions. But in practice, measured signals are usually discretized and represented in limited orders to reduce the complexity. These representations have adverse effects to the perturbation errors in measured signals, if the transformation matrices are ill-conditioned. For example, spherical harmonic representation of three dimensional sound fields involves representations in terms of spatially dependent continuous orthogonal functions. A discrete, limited order matrix representation called as Spherical Harmonic Matrix (SHM) is used for practical computations \cite{rafaely,Avni}. This representation can have adverse effects to the perturbation errors in measured signals, if the SHM is ill-conditioned \cite{rafaely,Abhaya_planewave}. So it is important to analyse the condition behaviour of SHM.

\footnote{\normalfont{This paper is a preprint of a paper submitted to IET Signal Processing Journal. If accepted, the copy of record will be available at the IET Digital Library. The first author of the paper is currently affiliated with center for vision, speech and signal processing, University of Surrey.}}
%


Condition number of any matrix determines the sensitivity of the output to a small change in the input under linear transformation. Condition number of a matrix is a non-convex and non-smooth function. General problem of minimizing the condition number of a matrix is a complex problem. However choosing a sub-matrix out of a rectangular matrix, with a subset of similar columns, that minimizes the condition number can be studied using optimization based techniques. Solving such an optimization problem has many applications in the domain of signal processing. Consider the spherical harmonic matrix as given below.

\vspace{-0.15cm}

\setcounter{MaxMatrixCols}{4}
\begin{equation}
\bold{Y}=\begin{bmatrix}
Y_{0}^{0}(\theta_1,\phi_1) & Y_{0}^{0}(\theta_2,\phi_2) & .. & Y_{0}^{0}(\theta_Q,\phi_Q) \\
 Y_{1}^{-1}(\theta_1,\phi_1) & Y_{1}^{-1}(\theta_2,\phi_2) & .. & Y_{1}^{-1}(\theta_Q,\phi_Q) \\
.& .& .& .\\
.& .& .& .\\
 Y_{N}^{N}(\theta_1,\phi_1) & Y_{N}^{N}(\theta_2,\phi_2)  & .. & Y_{N}^{N}(\theta_Q,\phi_Q) \\
\end{bmatrix}
\label{SHM}
\end{equation}
The behaviour of the matrix depends on the sampling positions chosen in the 3D space. It has been widely used in the domain of spherical array signal processing \cite{li2007flexible,Boaz_rafaely}, spatial audio \cite{jin_icassp,poletti2005three} etc. Particularly in the domain of spatial audio, sampling configurations are chosen for HRTF measurements \cite{Rodney} or loudspeaker positioning \cite{Abhaya_planewave,radmanesh2013generation,Pulkki}.  Minimizing the condition number of SHM can aid in choosing better sampling configurations that would prevent from being sensitive to the perturbation errors in the measurement process. But there exists many sampling configurations in the literature which are widely used in various domains. It is important to know the properties of each configuration so that their relevance in the domain of spatial audio can be analysed.

Sampling schemes around a sphere are widely studied in various domains \cite{mcewen2011novel,Gonz치lez2009,saff1997distributing,brauchart2015distributing}. 
Some of the well known spherical sampling schemes existing in the literature are T-designs of platonic solids \cite{Hardin1996}, Equiangle sampling, Gaussian sampling \cite{rafaely}, Fibonacci lattice \cite{Gonz치lez2009}, Fliege nodes \cite{fliege1999distribution}, Lebedev grids \cite{lecomte2016fifty},  Interaural sampling \cite{cipic} etc. Each of these designs have different properties which assumes high significance in particular application. For example, T-designs and Fliege nodes exhibit nearly uniform placement of points over a sphere and finds application in the loudspeaker positioning for spatial sound reproduction \cite{zotter2010virtual} and design of spherical microphone arrays \cite{li2007flexible,meyer2002highly}. A fifty-node Lebedev grids as discussed in \cite{lecomte2016fifty}, exhibits  better orthogonality among spherical harmonic vectors as compared to T-designs and Fliege nodes, and found to be important for recording and reproduction of sound fields. Equiangular sampling, Gaussian sampling are regular configurations which exhibit closed form expressions for angular directions  and quadrature weights. They have a practical advantage of having equal spacing between consecutive sampling points in the azimuthal plane. In fact, Equiangular sampling exhibits equal spacing between the sampling points in both azimuthal and elevation planes. But they suffer from dense sampling points at the poles. Fibonacci sampling exhibits nearly uniform sampling on a sphere with closed form expression for angular directions for any number of sampling points. Interaural sampling is considered to be important in the domain of spatial hearing where sampling points are chosen on interaural circles. In this manner, each sampling scheme has advantages in different applications. This work mainly focus on the condition behaviour of SHM for various sampling configurations that are relevant in the domain of spatial audio. In this regard, we mainly focus on two applications of spatial audio, namely, spatial sound reproduction using loudspeakers and HRTF measurement where choice of sampling configuration based on the conditioning of SHM assumes importance.

Sound field reproduction is the synthesis of a desired sound field over a particular region using multiple loudspeakers located in the 3D space. Many works have been earlier attempted to reproduce desired sound field by optimal choice of sampling configuration for loudspeakers \cite{Abhaya_planewave,radmanesh2013generation,asano1999optimization}. In \cite{Abhaya_planewave}, plane-wave reproduction using array of loudspeakers has been discussed where loudspeakers are treated as point sources. It also discuss the importance of the conditioning behaviour of SHM, and incorporates the T-design configurations  for loudspeaker positioning. Loudspeaker positioning for spatial field reproduction has also been studied as a power constrained error minimization problem \cite{khalilian2016comparison}. In \cite{jin_icassp}, sound field reconstruction using spherical harmonic representation called as Ambisonics has been discussed in a compressed sensing framework by assuming loudspeakers as plane wave sources. Among the above mentioned approaches, this work mainly focuses on the  Ambisonics way of reproducing sound fields.  In Ambisonics \cite{daniel1998ambisonics}, spatial sound is generally encoded in to spherical harmonics. These encoded signals are subsequently used to drive loudspeakers for the synthesis of spatial sound. The encoding and decoding process involves linear transformation using SHM. Choosing a proper configuration for loudspeakers is an important task to maintain a well conditioned SHM. T-designs were earlier used for positioning the loudspeakers in the 3D space  \cite{zotter2010virtual}. T-designs allow a uniform placement of loudspeakers in the 3D space, and also results in orthogonal spherical harmonic vectors with a well conditioned SHMs. But T-designs  are available only for a limited number of configurations, and for a maximum of 20 samples, supporting a maximum order of only $N = 2$ \cite{rafaely}.  T-designs for number of samples greater than 20 are  available only for few set of T-values and their corresponding spherical harmonic orders \cite{Hardin1996,neil_sloane}.  Hence the focus of this work is to identify configurations for any number of loudspeakers that can reproduce sound field efficiently and also result in well conditioned SHM. 

In the measurement of HRTFs, spatial configurations are chosen based on the following factors. The resolution of HRTF measurement is usually chosen based on the humans sensitivity to the spatialised sound perception. The configuration is generally chosen such that circular hoops can be utilised for simultaneous HRTF measurement. Earlier many HRTF sampling configurations like CIPIC \cite{cipic}, KEMAR \cite{kemar} were proposed considering the above factors.  But it is also important to choose spatial configuration which results in well conditioned SHM. It is so because, spherical harmonic representation of HRTFs plays an important role in the interpolation and range extrapolation of HRTFs \cite{durai_int_ext}. Additionally HRTFs of individual frequencies have been earlier expressed using SHM \cite{Avni,abhaya_insights} . So it is important to obtain a spatial configuration which results in well conditioned SHM and also considering the aforementioned factors.

The primary contributions of this work are as follows. A novel optimization based solution is developed for obtaining a sub-matrix of a rectangular matrix with a reduced condition number. Subsequently, the proposed method would be used to obtain a well conditioned SHM for spatial audio applications. Particularly spatial sampling configurations are identified for loudspeaker based spatial sound reproduction and HRTF measurement. The configurations obtained using the proposed method is evaluated for various spherical harmonic orders and compared with the existing configurations.

The rest of the paper is organized as follows. Section \ref{LAcond} describes the proposed method for minimizing the condition number of a rectangular sub-matrix using an optimization based approach. Section \ref{sec3} presents the identification of loudspeaker geometry and HRTF sampling configuration using the proposed method.  Section \ref{sec4} discusses the practical implementation, and the performance analysis of the proposed method.

\section{Condition Number Minimization of a Rectangular Sub-Matrix Using an Optimization based Approach}
\label{LAcond} 
Condition number of any matrix determines the sensitivity of the output to a small change in the input under linear transformation. In many applications, it is desirable to identify a sub matrix, by choosing a selected columns or rows of a original matrix such that, the sub-matrix would result in a lowest condition number. Such problems are difficult to handle by an exhaustive search as the computational complexity increases in a combinatorial way with the increase in the order of the matrix. Hence it is important to reformulate the problem such that the optimal solution can be identified for larger matrices with the available resources. An optimization based approach for solving above problem is discussed as follows.

\vspace{-0.2cm}

\subsection{Problem Formulation}
Consider a rectangular matrix $\mathbf{A}_{P\times Q}$ $(Q>>P)$. Condition number of $\mathbf{A}$, denoted as $\kappa(\mathbf{A})$, is defined as \cite{strang09}, 


\begin{equation}
\kappa(\mathbf{A})=||\mathbf{A}|| \hspace{0.1cm} ||\mathbf{A^{-1}}||
\end{equation}
where $||\cdot||$ is the operator norm. If the norms are defined in an Euclidean way, then the condition number of a matrix $\mathbf{A}$ can be computed as the ratio of maximum singular value to the minimum singular value of matrix $\mathbf{A}$. The expression for $\kappa(\mathbf{A})$ is given as,

\begin{equation}
\kappa(\mathbf{A})=\frac{\sigma_{max}(\mathbf{A})}{\sigma_{min}(\mathbf{A})}
\end{equation}
where $\sigma_{max}(\mathbf{A})$, $\sigma_{min}(\mathbf{A})$ represent maximum and minimum singular values of matrix $\mathbf{A}$ respectively. Any sub-matrix of $\mathbf{A}$ with dimension $P \times Q'$ can be constructed as

\begin{equation}
\mathbf{A'}=\mathbf{A}_{P\times Q}\mathbf{\Gamma}_{Q\times Q'}
\end{equation}
where $\mathbf{A}'$ is the sub-matrix with $Q'$ columns, and $\mathbf{\Gamma}_{Q\times Q'}$ is a rectangular binary matrix with $Q$ rows and $Q'$ columns $(Q>Q')$.  The objective of this work is to identify the optimal sub-matrix $\mathbf{A'}^*$ which has least condition number among all the sub-matrices of $\mathbf{A}$ with dimension $P\times Q'$. The objective function of the minimization problem is given as,

\begin{equation*}
\begin{aligned}
& {\text{min}}
& &  \kappa(\mathbf{A'}) & \equiv & & \underset{\mathbf{\Gamma}}{\text{min}} & & \kappa(\mathbf{A\Gamma})
\end{aligned}
\end{equation*}
The sub-matrix $\mathbf{A'}$ can contain a subset of columns of $\mathbf{A}$, only if, each column of $\mathbf{\Gamma}$ contains exactly one non-zero element and each row of $\mathbf{\Gamma}$ contains a maximum of one non-zero element. Imposing these constraints on $\mathbf{\Gamma}$ would result in a constrained optimization problem as given below.

\begin{equation}
\begin{aligned}
& \underset{\mathbf{\Gamma}}{\text{minimize}}
& &  \kappa(\mathbf{A\Gamma}) \\
& \text{subject to}
& & \gamma_{ij} \in {\{0,1\}} \\
& & &\sum_{i=1}^{Q} \gamma_{ij} =1 & \forall j \in \{1 \cdots Q'\}\\
& & &\sum_{j=1}^{Q'} \gamma_{ij} \leq 1 & \forall i \in \{1 \cdots Q\}\\
\end{aligned}
\label{opt1}
\end{equation}
where $\gamma_{ij}$ is an element of the $i^{th}$ row and $j^{th}$ column of matrix $ \mathbf{\Gamma}$. The problem posed in Equation \ref{opt1} is a constrained integer programming problem with binary variables. The objective function is non-convex, and hence it is re-formulated by utilizing the following properties of condition number.

\begin{equation}
\kappa(\mathbf{A})=\sqrt{\kappa(\mathbf{AA^{H}})} 
\end{equation}

\begin{equation}
\frac{\sigma_{max}(\mathbf{A})}{\sigma_{min}(\mathbf{A})}=\sqrt{\frac{\lambda_{max}(\mathbf{AA^{H}})}{\lambda_{min}(\mathbf{AA^{H}})}}
\label{property}
\end{equation}
where $\mathbf{H}$ is a Hermitian operator, $\lambda_{max}(\cdot)$ and $\lambda_{min}(\cdot)$ are the maximum and minimum eigenvalues of a corresponding matrix respectively. Using the property given in Equation \ref{property}, the objective function can be reformulated as,


\begin{eqnarray}
\kappa(\mathbf{A\Gamma})=\sqrt{\frac{\lambda_{max}(\mathbf{A\Gamma\Gamma^{H}A^{H}})}{\lambda_{min}(\mathbf{A\Gamma\Gamma^{H}A^{H}})}}=\sqrt{\frac{\lambda_{max}(\mathbf{A\Delta A^{H}})}{\lambda_{min}(\mathbf{A\Delta A^{H}})}} 
\label{diag}
\end{eqnarray}
where $\mathbf{\Delta}=\mathbf{\Gamma\Gamma^{H}}$, is a diagonal matrix with binary elements $(0,1)$. $\mathbf{\Delta}$ comprises of exactly $Q-Q'$ zeros as its diagonal elements.  The matrix $(\mathbf{A\Delta A^{H}})$ given in Equation \ref{diag} is a Hermitian positive definite matrix, if the sub-matrix $\mathbf{A\Gamma}$ is full row rank matrix. Otherwise $(\mathbf{A\Delta A^{H}})$ is a Hermitian positive semi-definite matrix whose minimum eigenvalue is zero and an infinite condition number. Therefore, it is important to maintain the matrix $\mathbf{A\Gamma}$ as full row rank. In practise, this is achieved by choosing $(Q'>>P)$. It is also a known fact that, the maximum and minimum eigenvalues of any Hermitian matrix are convex and concave functions respectively. Therefore the non-convex problem in Equation \ref{opt1} is reformulated as,

\begin{equation}
\begin{aligned}
\mathbf{\Delta^*}= \quad & \underset{\mathbf{\Delta, diagonal}}{\text{minimize}}
& & \lambda_{max}(\mathbf{A\Delta A^{H}}) \\
&\text{subject to}
& &\delta_{ii} \in {\{0,1\}} \quad  \forall i \in \{1 \cdots Q\}\\
& & &\sum_{i=1}^{Q} \delta_{ii} =Q' \\
& & &\lambda_{min}(\mathbf{A\Delta A^{H}}) > \eta^* 
\end{aligned}
\label{opt2}
\end{equation}

\begin{equation}
\kappa(\mathbf{A'}^{*})=\sqrt{\frac{\lambda_{max}(\mathbf{A\Delta^* A^{H}})}{\lambda_{min}(\mathbf{A\Delta^* A^{H}})}}
\label{condd}
\end{equation}
where $\delta_{ij}$ is an element of the $i^{th}$ row and $j^{th}$ column of matrix $ \mathbf{\Delta}$. $\mathbf{\Delta^*}$ is the optimal solution of the above constrained optimization problem for an optimal lower bound $\eta^*$ on the minimum eigenvalue of $\mathbf{A\Delta A^{H}}$. The proof for the equivalence of the optimization problems given in Equation \ref{opt1} and Equation \ref{opt2} is given in Appendix \ref{append1} . The optimum lower bound $\eta^*$ is an unknown quantity, but it lies in between an  upper and lower bound  as given below.

\begin{equation}
\eta_{lb}<=\eta^*<=\eta_{ub}.
\end{equation}
These bounds $\eta_{lb}$, and $\eta_{ub}$ are computed using the properties of eigen values of a Hermitian matrices. Its a known fact that, the eigenvalues of $\mathbf{A\Delta A^{H}}$ are positive. Hence the lower bound $\eta_{lb}$ can be considered as zero. The upper bound $\eta_{ub}$ is computed as,


\begin{equation}
\eta_{ub}= \frac{Tr (\mathbf{AA^{H}})}{P}.
\end{equation}
The derivation for upper bound is described in Appendix \ref{append2}. So the final bounds for $\eta^*$ is given as,


\begin{equation}
0 < \eta^* \leq \frac{Tr (\mathbf{AA^{H}}) }{P}.
\label{LBUB1}
\end{equation}
The computation of optimal lower bound $(\eta^*)$ from the above given range is an important task of the proposed solution. A complete methodology to identify the optimal lower bound is discussed as follows.

\begin{figure*}[t]


\newcommand{\midarrow}{\tikz \draw[-triangle 90](0,0) -- +(.1,0);}
\begin{tikzpicture}[node distance=2.4cm]
\tikzstyle{arrow} = [thick,->,>=stealth]
\tikzstyle{arrow1} = [thick,<-,>=stealth]
\tikzstyle{arrow2} = [thick,-,>=stealth]
\tikzstyle{arrow3} = [dashed,-,>=stealth]

\node at (0.6,0) {}; 
  
   \draw decorate [decoration={brace,mirror,amplitude=4pt},xshift=2pt,yshift=0pt] {(4,-0.3) -- (6.8,-0.3)};
         \draw decorate [decoration={brace,mirror,amplitude=4pt},xshift=2pt,yshift=0pt] {(6.8,-0.3) -- (9.5,-0.3)};
      \draw decorate [decoration={brace,mirror,amplitude=4pt},xshift=2pt,yshift=0pt] {(13,-0.3) -- (15.5,-0.3)}; 
 
\draw (4,0.4)--(4,0.8);   
\draw (6.8,0.4)--(6.8,0.8); 
\draw (9.5,0.4)--(9.5,0.8); 
\draw (13,0.4)--(13,0.8); 
\draw (15.5,0.4)--(15.5,0.8); 

\node at (1.5,1.2) {Lower Bound $(\eta) \hspace{0.7cm} :$};
 \node at (4,1.2) {$\eta^0=0$};
  \node at (6.8,1.2) {$\eta^1=\lambda_{min}^{0}$};
   \node at (9.5,1.2) {$\eta^2=\lambda_{min}^{1}$};
    \node at (13.5,1.2) {$\eta^{R-1}=\lambda_{min}^{R-2}$};
  \node at (15.5,1.2) {$\frac{Tr(\mathbf{AA^H})}{P}$};


  \node at (1.5,-1.3) {Diagonal Matrix $(\mathbf{\Delta}) \hspace{0.3cm} :$};
   \node at (1.5,-2) {Sub-matrix $(\mathbf{A\Delta A^H})\hspace{0.3cm} :$};
    \node at (1.5,-2.8) {Maximum Eigenvalue \hspace{0.3cm} : };
    \node at (1.5,-3.2) {of Sub-matrix $(\lambda_{max}) \hspace{0.4cm} $};
    \node at (1.5,-4) {Minimum Eigenvalue \hspace{0.3cm} :};
    \node at (1.5,-4.4) { of Sub-matrix $(\lambda_{min}) \hspace{0.4cm} $};
     \node at (1.5,-5.2) {Condition Number $(\kappa) \hspace{0.2cm} :$};

  \node at (5.5,-1.3) {$\mathbf{\Delta}^0$};
   \node at (5.5,-2) {$\mathbf{A\Delta}^0\mathbf{A^H}$};
    \node at (5.5,-2.8) {$\lambda_{max}^{0}$};
     \node at (5.5,-4) {$\lambda_{min}^{0}$};
     \node at (5.5,-5.2) {$\kappa^{0}$};

  \node at (8.1,-1.3) {$\mathbf{\Delta}^1$};
   \node at (8.1,-2) {$\mathbf{A\Delta}^1\mathbf{A^H}$};
    \node at (8.1,-2.8) {$\lambda_{max}^{1}$};
     \node at (8.1,-4) {$\lambda_{min}^{1}$};
     \node at (8.1,-5.2) {$\kappa^{1}$};
     

  \node at (11.5,-1.3) {$\cdots$};
   \node at (11.5,-2) {$\cdots$};
    \node at (11.5,-2.8) {$\cdots$};
     \node at (11.5,-4) {$\cdots$};
     \node at (11.5,-5.2) {$\cdots$};

  \node at (14.5,-1.3) {$\mathbf{\Delta}^{R-1}$};
   \node at (14.5,-2) {$\mathbf{A\Delta}^{R-1}\mathbf{A^H}$};
    \node at (14.5,-2.8) {$\lambda_{max}^{R-1}$};
     \node at (14.5,-4) {$\lambda_{min}^{R-1}$};
     \node at (14.5,-5.2) {$\kappa^{R-1}$};
     




  \draw  (4,0.6)--(7,0.6);
  \draw  (7,0.6)--(9.5,0.6);
  \draw [arrow3] (9.5,0.6)--(13,0.6);
   \draw  (13,0.6)--(15.5,0.6);
 
\end{tikzpicture}
\vspace{0.4cm}
\caption{Figure illustrating $R$ transitions of lower bound $\eta$ and the corresponding optimum solution $(\mathbf{\Delta})$ resulted at each transition by the optmization problem.}
\label{Flowdia}
\end{figure*}

\subsection{Computation of Optimal Lower Bound of Minimum Eigenvalue}
\label{optLB}

Consider a variable $\eta$ denoting a lower bound value in the range $[\eta_{lb},\eta_{ub}]$. The objective of the work is to find an optimum value of $\eta$ in its range, denoted as $\eta^*$, for which condition number of $\mathbf{A\Delta A^{H}}$ is minimum. This is expressed as given below.

 
  \begin{equation}
 \begin{aligned}
\mathbf{\eta}^* =\min_{\eta}\{\kappa(\mathbf{A\Delta A^{H}})\} \quad 0\leq \eta \leq \frac{Tr (\mathbf{AA^{H}})}{P}.
\end{aligned}
 \end{equation}
However the range of $\eta$ values are continuous. But only a discrete and finite number of $\eta$ values has actual significance in the computation of optimal lower bound $\eta^*$. Because, for over a range of $\eta$ values, the optimization problem given in Equation \ref{opt2} will result in a similar  binary diagonal matrix $\mathbf{\Delta^*}$, and a similar condition number. In order to better understand this peculiar behaviour, few additional terms  are introduced and computation of lower bounds at the transition points is also presented. The value of $\eta$ where there is a transition in the value of condition number is computed as,

\begin{equation}
\eta^{i}=\lambda_{min}^{i-1} \quad \quad i=1 \cdots R
\label{stepeta}
\end{equation}
where $\eta^{i}$ is the lower bound of minimum eigenvalue at the $i^{th}$ transition point, and $\lambda_{min}^{i-1}$ is the minimum eigenvalue of the optimum sub-matrix $(\mathbf{A\Delta}^{i-1} \mathbf{A^{H}})$ resulted in the $(i-1)^{th}$ transition. The binary diagonal matrix resulted for the lower bound $\eta^i$ is computed as,


\begin{equation}
\begin{aligned}
\mathbf{\Delta}^i= \quad & \underset{\mathbf{\Delta, diagonal}}{\text{minimize}}
& & \lambda_{max}(\mathbf{A\Delta A^{H}}) \\
&\text{subject to}
& &\delta_{ii} \in {\{0,1\}} \quad \forall i \in \{1 \cdots Q\}\\
& & &\sum_{i=1}^{Q} \delta_{ii} =Q' \\
& & &\lambda_{min}(\mathbf{A\Delta A^{H}}) > \eta^i 
\end{aligned}
\label{optdiscrete}
\end{equation}
Equation \ref{optdiscrete} is solved using numerical techniques \cite{Lofberg2004}. Implementation issues are discussed in section \ref{prac_imple}. The sub-matrix $(\mathbf{A\Delta}^i\mathbf{A^H})$, its minimum eigenvalue $\lambda_{min}^{i}$, maximum eigenvalue $\lambda_{max}^{i}$, and the condition number $\kappa^{i}$ for $R$ intervals is illustrated in Figure \ref{Flowdia}. The reason for an optimization problem to result in a similar solution for values of $\eta$ lying between $(\eta^{i},\eta^{i+1})$ is described as follows.

\begin{figure}[h]

\vspace{-0.3cm}

\newcommand{\midarrow}{\tikz \draw[-triangle 90](0,0) -- +(.1,0);}
\begin{tikzpicture}[node distance=2.4cm]
\tikzstyle{arrow} = [thick,->,>=stealth]
\tikzstyle{arrow1} = [thick,<-,>=stealth]
\tikzstyle{arrow2} = [thick,-,>=stealth]
\tikzstyle{arrow3} = [dashed,-,>=stealth]

\node at (0.6,0) {};

\draw (1,-0.4)--(1,0.4);   
\draw (3,-0.4)--(3,0.4); 
\draw (4.6,-0.4)--(4.6,0.4); 
\draw (8,-0.4)--(8,0.4);

 \node at (1,0.8) {$\eta^i=\lambda^{i-1}_{min}$};
 \node at (3,0.8) {$\hat{\eta}$};
 \node at (4.6,0.8) {$\eta^{i+1}=\lambda^{i}_{min}$};
  \node at (7.6,0.8) {$\eta^R=\frac{Tr(\mathbf{AA^H})}{P}$};
  
  \node at (1.2,-0.6) {$\rightarrow$};
   \node at (1.2,-0.9) {$\lambda_{min}(\cdot)>\eta^i$};
     \node at (3.2,-0.6) {$\rightarrow$};
   \node at (3.2,-0.9) {$\lambda_{min}(\cdot)>\hat{\eta}$};

  \draw  (1,0)--(5.5,0);
  \draw [arrow3] (5.6,0)--(6.8,0);
   \draw  (6.8,0)--(8,0);
 
\end{tikzpicture}
\vspace{-0.3cm}
\caption{Figure illustrating the feasible regions of $\lambda_{min}(\mathbf{A\Delta A^{H}})$ function for lower bound $\hat{\eta}$ and lower bound of $i^{th}$ transition $\eta^i$. The optimal solution at the $i^{th}$ transition $(\lambda^{i}_{min})$  is in the feasible region of $\lambda_{min}(\cdot)>\hat{\eta}$.}
\label{linedia1}
\end{figure}
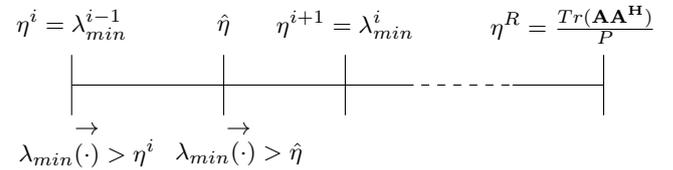

Consider $\eta=\hat{\eta}$, where $\eta^{i}<\hat{\eta}<\eta^{i+1}$. The feasible region for $\lambda_{min}(\mathbf{A\Delta A^{H}})>\hat{\eta}$ is a subset of the feasible region for $\lambda_{min}(\mathbf{A\Delta A^{H}})>\eta^{i}$ as illustrated in Figure \ref{linedia1}. Additionally, the minimum eigenvalue of optimum sub-matrix at $i^{th}$ transition is greater than $\hat{\eta}$ as given below.
 
 
 \begin{equation}
   \lambda_{min}(\mathbf{A\Delta}^i \mathbf{A^{H}}) =\lambda_{min}^i>\hat{\eta} 
 \end{equation}
Unless the feasible region for lower bound $\hat{\eta}$ excludes the optimum solution generated at the $i^{th}$ transition, i.e $\hat{\eta}>\lambda_{min}^i$, the optimal solution for lower bound $\hat{\eta}$ will always be same as lower bound $\eta^{i}$. This can be mathematically given as follows.

  \begin{equation}
   \hat{\mathbf{\Delta}} =\mathbf{\Delta}^i, \quad \lambda_{min}^i=\hat{\lambda}_{min}, \quad  \kappa^i=\hat{\kappa} 
 \end{equation}
where $\hat{\lambda}_{min}$, $\hat{\kappa}$ are the minimum eigenvalue and condition number of sub-matrix $\mathbf{A\hat{\Delta}} \mathbf{A^{H}}$. Hence there can only be finite number $(R)$ of distinct diagonal matrices $\mathbf{\Delta}^i$ and condition numbers $\kappa^i$. Theoretically $R$ can take any  value less than ${Q}\choose{Q'}$. So the practical feasibility of the proposed algorithm cannot be generalised for all the rectangular matrices. However the current works restrict to Spherical Harmonic Matrices (SHM) which is found be exhibiting  fewer number of transitions $(R)$ which will be discussed in the section \ref{condfibo}. Total number of transitions $(R)$ would be smaller if the step size is larger which in turn depends on the minimum eigenvalue generated at that particular transition. The steps involved in identifying the $R$ lower bounds, and its corresponding $R$ diagonal matrices and $R$ condition numbers is enumerated in Algorithm \ref{alg1}. Hence the optimal lower bound for minimum eigenvalue is computed as

\begin{equation}
\begin{aligned}
\mathbf{\eta}^* =\arg\{\min\{\kappa_i(\eta^i)\} \} \quad i \in (0,R-1)
\end{aligned}
\label{opti_eta}
\end{equation}
The optimal $\mathbf{\Delta}^*$ and its corresponding condition number can now be computed as

 \begin{equation}
\begin{aligned}
\mathbf{\Delta}^*= \quad & \underset{\mathbf{\Delta, diagonal}}{\text{minimize}}
& & \lambda_{max}(\mathbf{A\Delta A^{H}}) \\
&\text{subject to}
& &\delta_{ii} \in {\{0,1\}} \quad  \forall i \in \{1 \cdots Q\}\\
& & &\sum_{i=1}^{Q} \delta_{ii} =Q' \\
& & &\lambda_{min}(\mathbf{A\Delta A^{H}}) > \eta^*
\end{aligned}
\label{optfinal}
\end{equation}

\begin{equation}
\kappa(A'^*)=\sqrt{\frac{\lambda_{max}(\mathbf{A\Delta}^* \mathbf{A^{H}})}{\lambda_{min}(\mathbf{A\Delta}^* \mathbf{A^{H}})}}
\label{condd1}
\end{equation}
where $\mathbf{\Delta}^*$ is the optimal binary diagonal matrix resulted from Equation \ref{optfinal} and $\kappa(A'^*)$ is the minimum condition number obtained for the lower bound $\eta^*$. An example based illustration of the proposed method is discussed in Appendix \ref{append3}.

\begin{algorithm}[h]
\caption{Algorithm to identify $R$ distinct lower bounds, and its corresponding $R$ diagonal matrices and $R$ condition numbers using the proposed method}
\label{alg1}
\begin{algorithmic}[1]
\State {\bf Input:} $\eta=0$, $i=0$, $Q'$, $\mathbf{A}$.

\While{$\eta < \frac{Tr (\mathbf{AA^{H}})}{P}$}
\State $\eta^{i}=\eta$
\State The optimal $\mathbf{\Delta}^i$ for lower bound $\eta^i$ is obtained by solving Equation \ref{optdiscrete} using optimization toolboxes as discussed in section \ref{prac_imple}.
\State $\lambda^{i}_{min}=\lambda_{min}(\mathbf{A\Delta}^i \mathbf{A^{H}})$
\State $\kappa^i=\sqrt{\frac{\lambda_{max}(\mathbf{A\Delta}^i \mathbf{A^{H}})}{\lambda_{min}(\mathbf{A\Delta}^i \mathbf{A^{H}})}}$

\State  $\eta=\lambda^{i}_{min}$
\State $i \gets i+1$, 
\EndWhile
\State Total number of subsets $R=i$.
\State {\bf Output:} $R$ distinct lower bounds $\eta^0 \cdots \eta^{R-1}$, and its corresponding diagonal matrices $(\mathbf{\Delta}^0, \mathbf{\Delta}^1, \cdots \mathbf{\Delta}^{R-1})$ and  condition numbers $(\kappa^0, \kappa^1, \cdots \kappa^{R-1})$.  
\end{algorithmic}
\end{algorithm}

\begin{figure*}[t]
\hspace{1cm}
\begin{minipage}[h]{0.166\linewidth}
  \centering
  \centerline{\includegraphics[width=3.5cm,height=3.5cm,trim=450 50 350 0,clip]{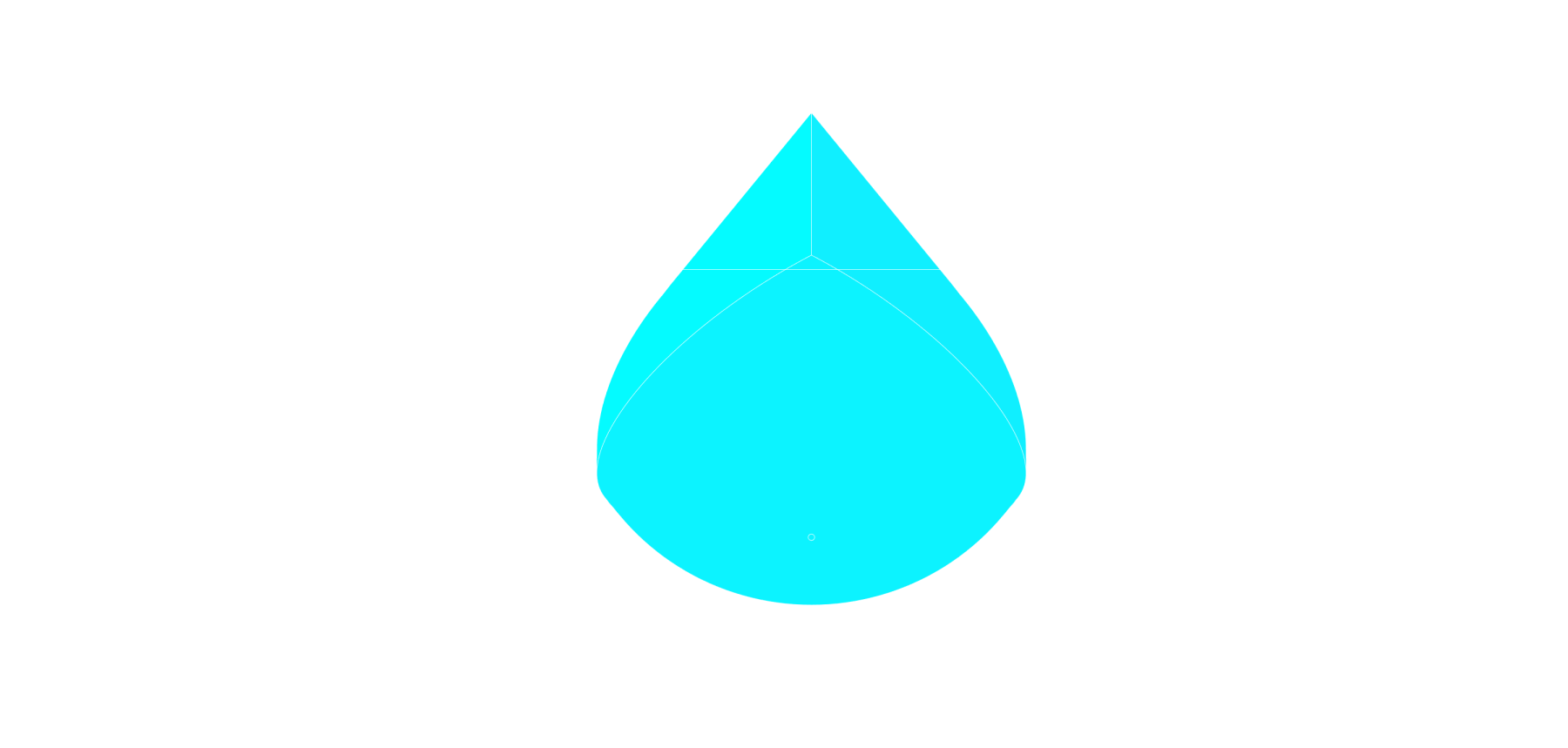}}
  \vspace{-0.3cm}
  \centerline{\footnotesize T=2, Q=4}\medskip
   \centerline{$(a)$}\medskip
\end{minipage}
\begin{minipage}[h]{0.166\linewidth}
  \centering
  \centerline{\includegraphics[width=3.5cm,height=3.5cm,trim=350 100 350 80,clip]{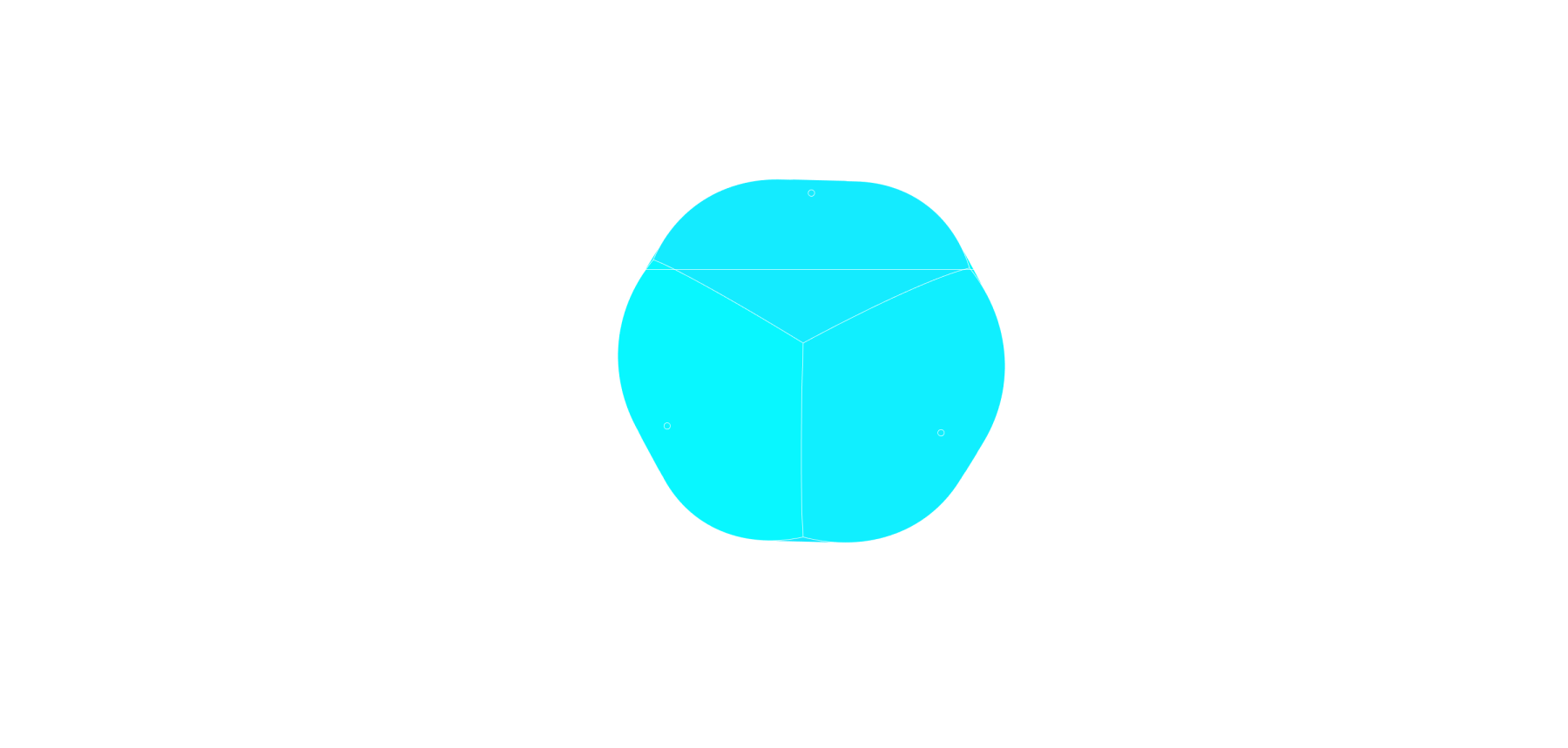}}
  \vspace{-0.3cm}
  \centerline{\footnotesize T=3, Q=6}\medskip
  \centerline{$(b)$}\medskip
\end{minipage}
\begin{minipage}[h]{0.166\linewidth}
  \centering
  \centerline{\includegraphics[width=3.5cm,height=3.5cm,trim=350 60 300 100,clip]{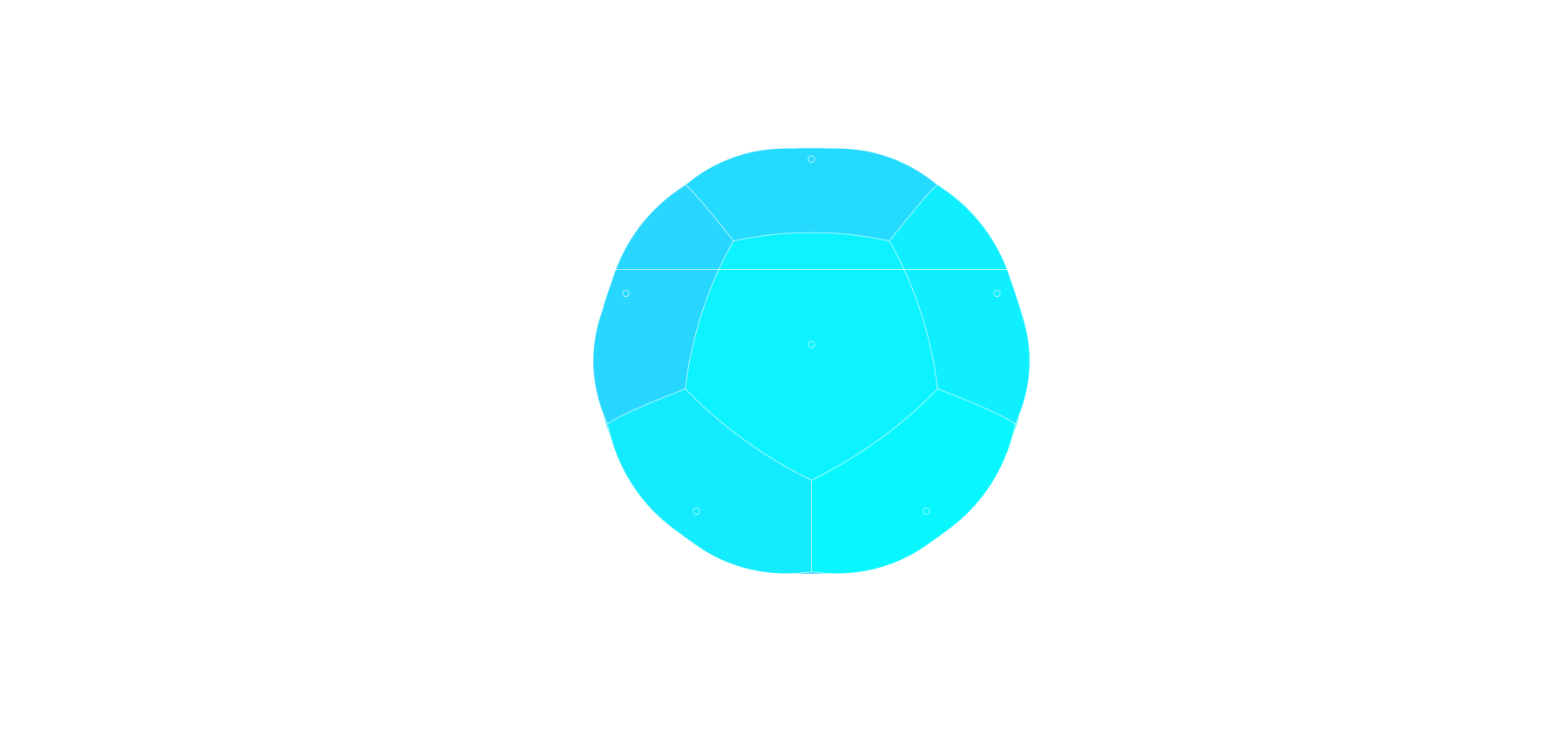}}
  \vspace{-0.3cm}
  \centerline{\footnotesize T=5, Q=12}\medskip
  \centerline{$(c)$}\medskip
\end{minipage}
\begin{minipage}[h]{0.166\linewidth}
  \centering
  \centerline{\includegraphics[width=3.5cm,height=3.5cm,trim=300 0 300 0,clip]{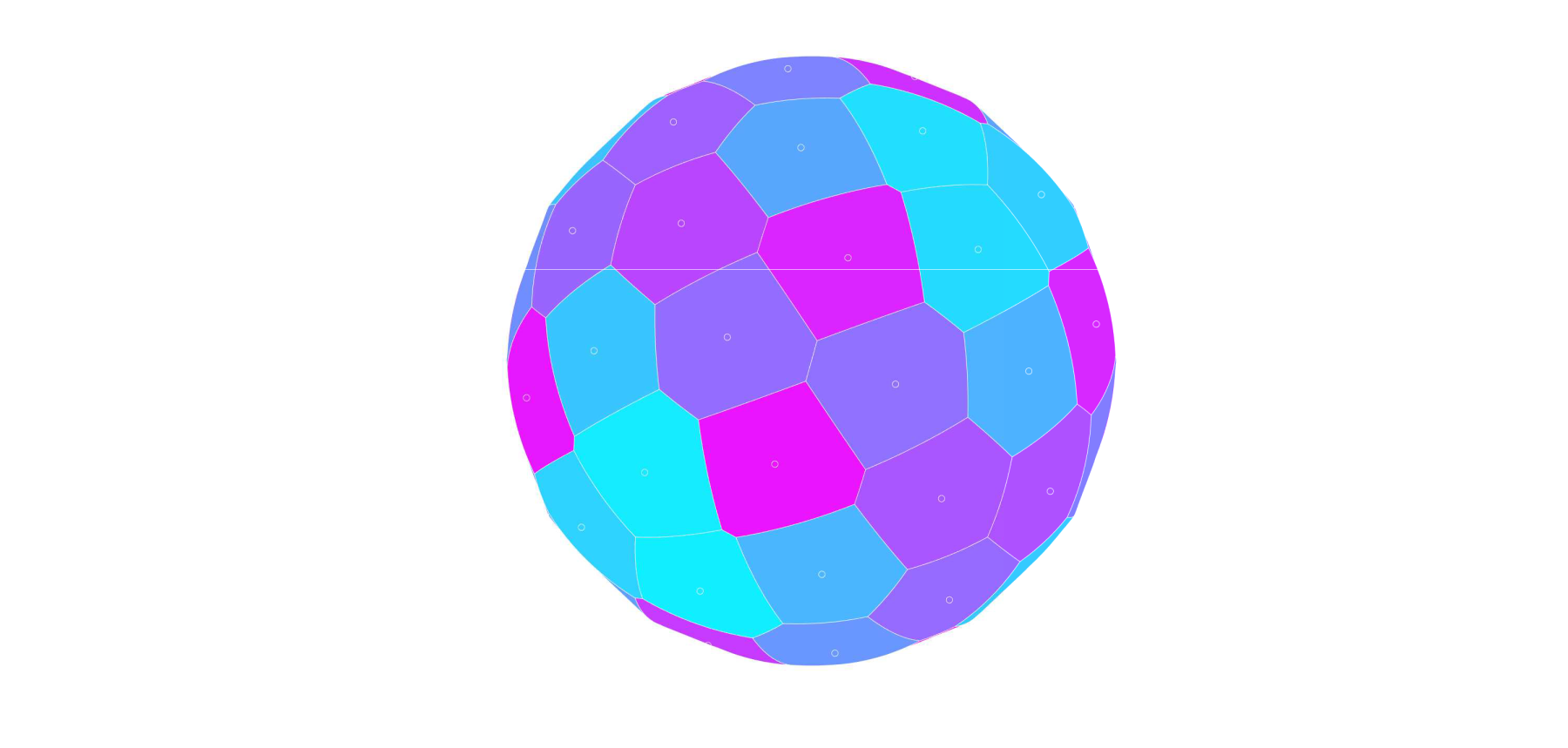}}
  \vspace{-0.4cm}
  \centerline{\footnotesize T=10, Q=60}\medskip
  \centerline{$(d)$}\medskip
\end{minipage}
\begin{minipage}[h]{0.17\linewidth}
  \centering
  \centerline{\includegraphics[width=3.5cm,height=3.5cm,trim=300 0 400 0,clip]{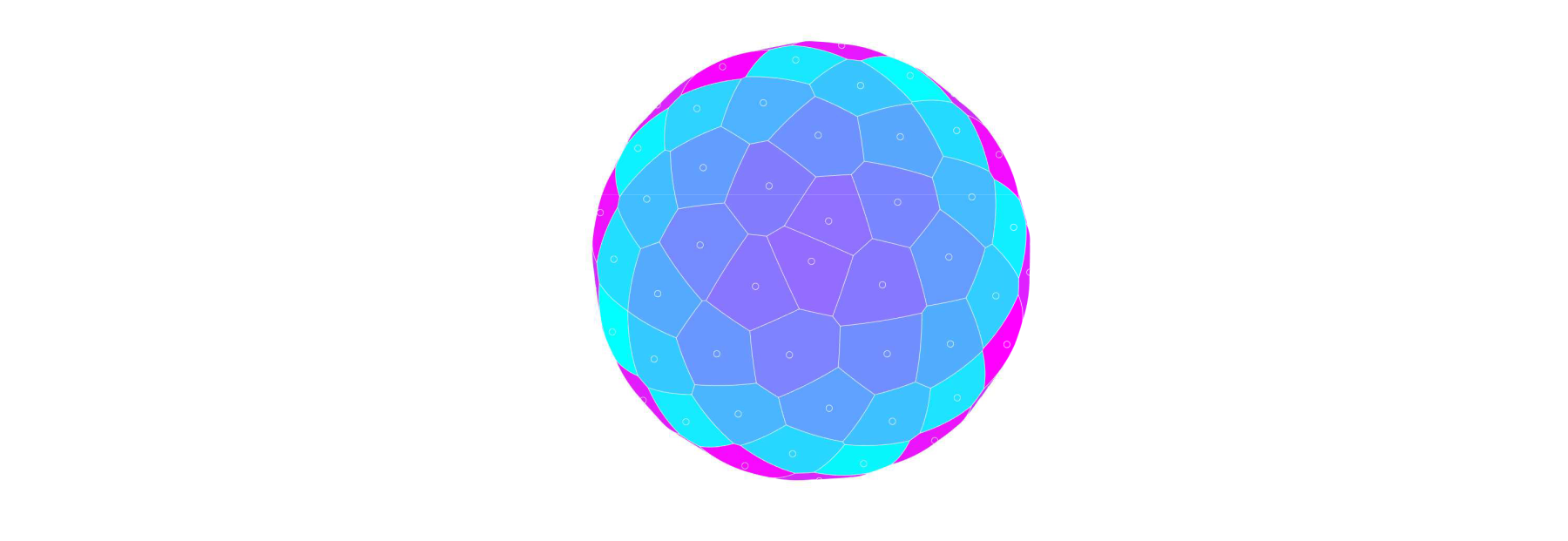}}
  \vspace{-0.4cm}
  \centerline{\hspace{0.4cm} \footnotesize Fibonacci Q=100}\medskip
  \centerline{\hspace{0.4cm} $(e)$}\medskip
\end{minipage}
\vspace{-0.25cm}
\caption{Figures (a-d) illustrates the voronoi polygonal diagrams of T-designs (T) for various number of sampling points $(Q)$. Figure (e) illustrates the voronoi polygonal diagram for $Q=100$ following Fibonacci lattice.}
\label{fig:voronoi_tdesign}
\end{figure*}

\section{Applications}
\label{sec3}
A  method to minimize the condition number of a rectangular sub-matrix was presented in Section \ref{LAcond}. In the current section, the proposed method together with application specific constraints is used in solving two different problems that arise in the domain of spatial audio. These problems include the identification of loudspeaker geometry for spatial sound reproduction, and the identification of spatial sampling configurations for HRTF measurement.  Both these problems and the proposed solutions are discussed in a detailed manner.



\subsection{Identification of Loudspeaker Geometry for Spatial Sound Reproduction}
\label{LSgeometry}

Loudspeaker geometry for spatial sound reproduction is chosen mostly based on the practical feasibility of loudspeaker positioning in the 3D space, ability to reproduce the desired sound field, and the conditioning behaviour of the SHM. Although T-designs of platonic solids were quiet apt in satisfying these factors, but they are limited to a maximum sampling positions of $Q=20$. T-designs  for number of samples $Q>20$ exhibiting nearly uniform configuration are available for only few set of values  as described in \cite{Hardin1996},\cite{neil_sloane}. Lack of uniform configurations for any possible number of sampling points and spherical harmonic orders motivates us to explore other configurations which maintain closer properties as T-designs. In this section, we firstly introduce a measure to quantify the uniformness of sampling points on sphere. Subsequently the importance of SHM in loudspeaker geometry identification is discussed. Further minimization of condition number of SHM using the proposed method method would be discussed for identifying the best geometry of loudspeakers for positioning in the 3D space.

\subsubsection{Measure of Uniformity for Spatial Distribution of Sampling Points on a Sphere}
\label{uni_D_measure}
The uniformity  of sampling points distributed on a sphere has been earlier measured using the voronoi density, which is inverse of the voronoi polygonal area \cite{bauer2000distribution}. The variation of these densities across all the polygons can be utilised as a measure  for uniformity of sampling points on the sphere. In \cite{bauer2000distribution}, this measure is called as D-measure which is given as,

\vspace{-0.5cm}

\begin{figure}
\hspace{0.3cm}
  \centerline{\includegraphics[width=10cm,height=5cm,trim=400 180 300 150,clip]{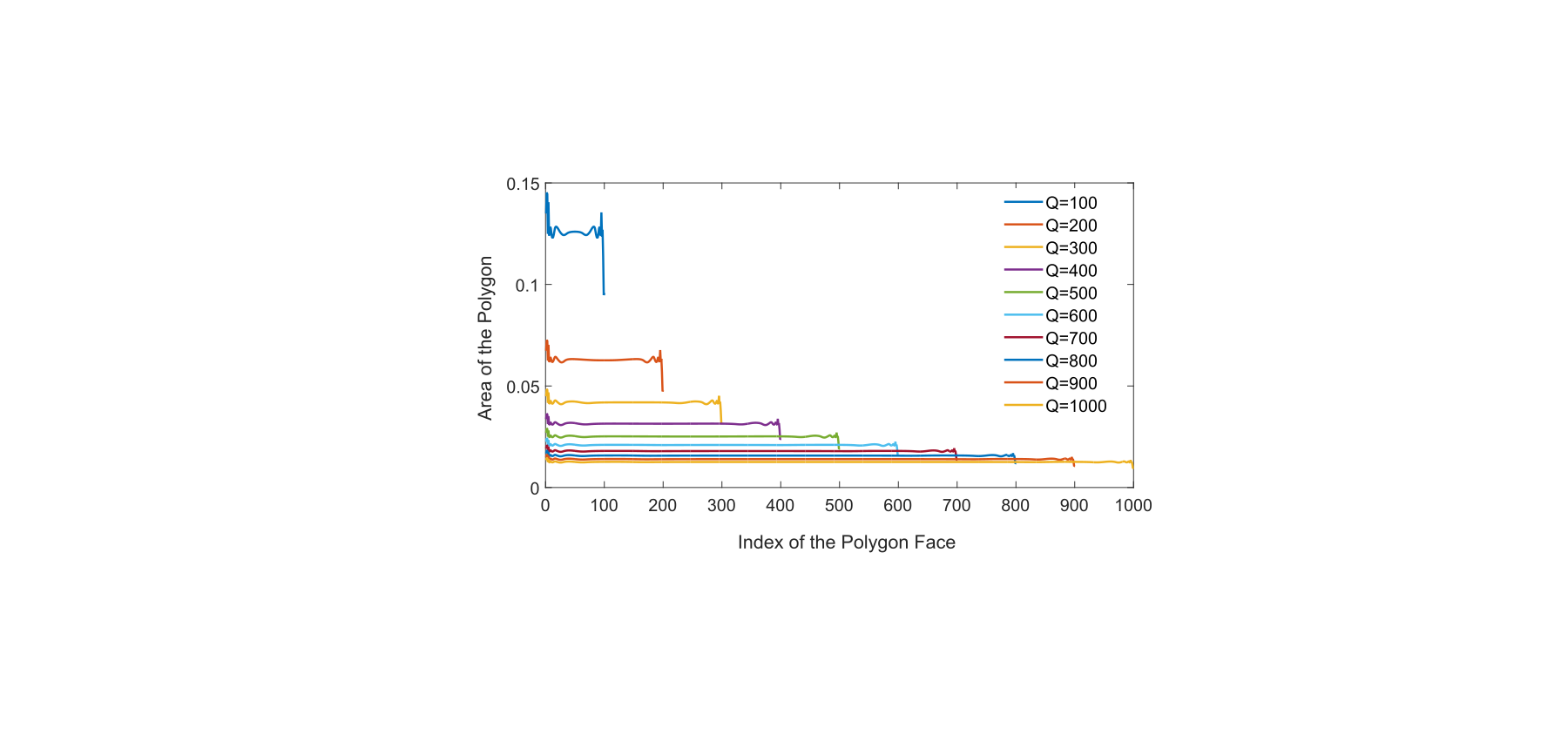}}
  \caption{Figure illustrating the variation of polygonal area for various faces of the polygon centred by sampling points generated using Fibonacci lattice.}
  \label{fig:Fiboarea}
\end{figure}

\vspace{-0.2cm}

\begin{eqnarray}
\nu&=&\sum_{k=1}^{Q}\frac{a_k}{4\pi} (d_k-\hat{d})^2 \\
D&=&\frac{\nu \sqrt(2\pi)}{\hat{d}}
\end{eqnarray}
where $a_k$ is the area of the $k^{th}$ voronoi polygon. $d_k=\frac{1}{a_k}$ is the density of the $k^{th}$ voronoi polygon. $\hat{d}=\frac{Q}{4\pi}$ is the average density over the sphere.  The voronoi polygons formed for $(T=2, Q=4)$, $(T=3,Q=6)$, $(T=5, Q=12)$, and $(T=10, Q=60)$ are illustrated in Figure \ref{fig:voronoi_tdesign}(a-d).  It can be observed that, all the polygons of any particular T-design exhibit equal areas, and thus an equal voronoi densities.  Hence the D-measure for T-designs is zero. However T-designs are  available only for limited  number of sampling points and spherical harmonic orders. But it is important to identify optimal geometry for loudspeakers for any number of sampling points. So a nearly uniform configuration of sampling points on sphere called as Fibonacci lattice \cite{Gonz치lez2009} is used as the reference for identifying the optimal loudspeaker geometry. The voronoi polygonal plot for Fibonacci lattice is given in Figure \ref{fig:voronoi_tdesign}e of 100 sampling points on the sphere. The variance of the polygonal area for various number of sampling points and polygonal faces is given in Figure \ref{fig:Fiboarea}. It can be noted that, except for few faces the polygonal area is nearly equal for almost all polygonal faces across various number of sampling points. But the condition number of SHM for Fibonacci lattice is high. For example, condition number of SHM with $Q=32, N=3$ is 1670. So the conditioning behaviour of SHM for Fibonacci lattice is minimized by appropriate selection of sampling points. The ensuing section first discusses the importance of SHM in loudspeaker positioning. Subsequently a method to identify the sampling points that results in a well conditioning behaviour of SHM and efficient reproduction of spatial sound is discussed.

\subsubsection{Importance of SHM for Identification of Loudspeaker Geometry}
\label{SHM_pw_LS}
Spherical harmonic functions has been widely used in the domain of spatial audio for 3D sound field reproduction using loudspeakers. Particularly in the techniques like Higher Order Ambisonics (HOA), spherical harmonic analysis is considered to be a powerful tool for describing the spatial properties of sound fields. Sound field emitted by a set of ideal planewave sources can be fully described by a set of spherical harmonic components as \cite{jin_icassp},


\begin{eqnarray}
\mathbf{\tilde{b}}(t)=\mathbf{\tilde{Y}}\mathbf{s}(t) \hspace{0.5cm} 
\label{repro_pw}
\end{eqnarray}

\vspace{-0.5cm}

\setcounter{MaxMatrixCols}{4}
\begin{equation}
\bold{\tilde{Y}}=\begin{bmatrix}
Y_{0}^{0}(\tilde{\theta}_1,\tilde{\phi}_1) & Y_{0}^{0}(\tilde{\theta}_2,\tilde{\phi}_2) & .. & Y_{0}^{0}(\tilde{\theta}_P,\tilde{\phi}_P) \\
 Y_{1}^{-1}(\tilde{\theta}_1,\tilde{\phi}_1) & Y_{1}^{-1}(\tilde{\theta}_2,\tilde{\phi}_2) & .. & Y_{1}^{-1}(\tilde{\theta}_P,\tilde{\phi}_P) \\
.& .& .& .\\
.& .& .& .\\
 Y_{N}^{N}(\tilde{\theta}_1,\tilde{\phi}_1) & Y_{N}^{N}(\tilde{\theta}_2,\tilde{\phi}_2)  & .. & Y_{N}^{N}(\tilde{\theta}_P,\tilde{\phi}_P) \\
\end{bmatrix}
\label{SHM1}
\end{equation}
where $\mathbf{s}(t)$ is the signal vector of $P$ plane wave sources, $\mathbf{\tilde{Y}}$ is the SHM truncated to order N,  and $\mathbf{\tilde{b}}(t)$ is the vector of spherical harmonic coefficients. It has to be noted that, the angular convention for elevation is measured from the z-axis. The sound field produced by $P$ plane wave sources is reproduced by a set of $Q$ loudspeakers $(Q>P)$ as given below:

\vspace{-0.5cm}

\begin{eqnarray}
\mathbf{b}(t)=\mathbf{Y}\mathbf{g}(t), \hspace{0.5cm} 
\label{encode}
\end{eqnarray}
where $\mathbf{Y}$ is SHM of Q sources as given earlier in Equation \ref{SHM}, $\mathbf{g(t)}$ indicates the set of loudspeaker signals which are assumed to be emitting plane-waves herein \cite{jin_icassp}. $\mathbf{b(t)}$ is the reproduced sound field coefficients. It has to be noted that, $Q$ directions of loudspeakers are different as compared to the $P$ directions of  plane wave sources. Equation  \ref {encode} is usually termed as the encoding process. The equivalent decoding process is to generate the driving signals for loudspeakers  using the spherical harmonic coefficients to synthesize 3D sound fields. Decoding process is given as,

\begin{equation}
\mathbf{g(t)=D\tilde{b}(t)},
\label{deco1}
\end{equation}
where $\mathbf{D}$ is the decoding matrix. Two widely used  methods  to compute the decoding matrix $\mathbf{D}$ are given as follows \cite{zotter2010virtual, poletti2005three}.\\

\begin{figure*}[b!]
  \centerline{\includegraphics[width=23cm,height=4.5cm,trim=100 322 40 152,clip]{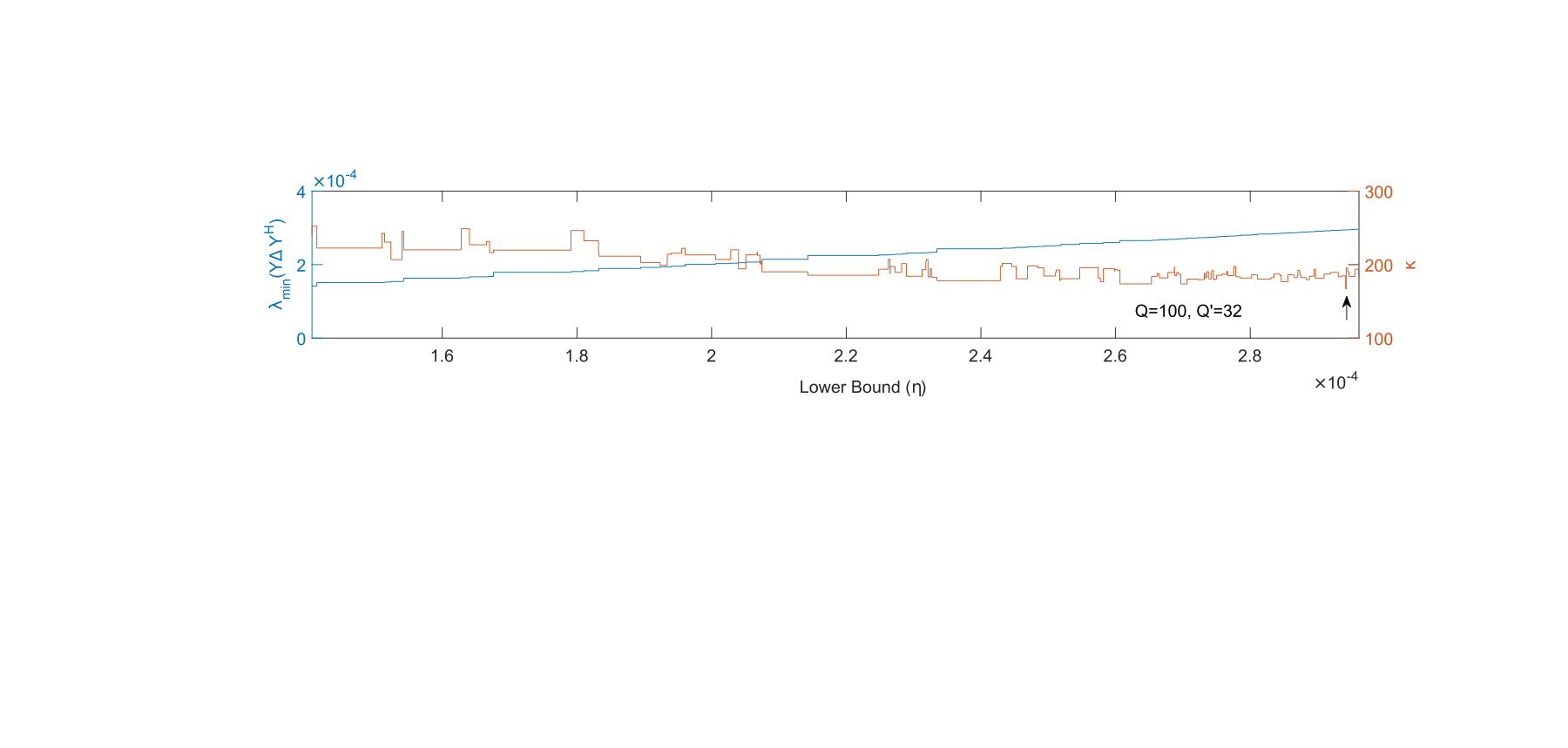}}
  \vspace{-0.2cm}
  \caption{Figure illustrating minimum eigenvalue and condition number of the sub-matrix $(\mathbf{Y\Delta}\mathbf{Y^{H}})$ for various lower bounds $(\eta)$ for selecting $Q'=32$ optimal sampling points from a total number of sampling points of $Q=100$ . }
  \label{fig:mineig_cond}
\end{figure*}

1. Sampling Decoder:
\vspace{-0.3cm}

\begin{equation}
\mathbf{D}=\frac{4\pi}{Q} \mathbf{Y^T}
\label{SD}
\end{equation}

2. Modematching Decoder:
\begin{equation}
\mathbf{D}=\mathbf{Y^T}(\mathbf{YY^T})^{-1}
\label{MMD}
\end{equation}
It has to be noted that, in both the encoding and decoding process, SHM is used as a transformation matrix. Particularly in the decoding process SHM is used in its transpose and least square form. A well conditioned decoding matrix will prevent from being sensitive to the perturbation errors. But it is known that,


\begin{equation}
\kappa(\mathbf{Y}^T) =   \kappa(\mathbf{Y}),
\end{equation}



\begin{table*}[b!]
\centering
\caption{Table depicts the total number of transition points $(R)$, optimal lower bound $(\eta^*)$, condition number $(\kappa)$ for selecting $Q'=32$ optimal sampling points from a set of $(Q)$ spatial sampling points.}
\label{table1}
\vspace{0.1cm}
\begin{tabular}{cccccccccccc}
\hline \hline \\
                     & Q=50   & Q=55   & Q=60   & Q=65   & Q=70   & Q=75   & Q=80   & Q=85   & Q=90   & Q=95   & Q=100  \\ \hline \\
$\eta^*(10^{-4})$ & 1.0851 & 1.1996 & 1.6513 & 1.3332 & 2.0353 & 1.4502 & 1.8704 & 2.2368 & 2.6939 & 2.2368 & 2.5471 \\  \hline \\
$\kappa$             & 289.38 & 257.74 & 213.48 & 232.58 & 182.21 & 229.37 & 209.96 & 189.38 & 147.56 & 174.65 & 166.73 \\  \hline \\ 
R                    & 63     & 59     & 134    & 68     & 76     & 6      & 74     & 100    & 49     & 23     & 133 \\  \hline \hline
\end{tabular}
\end{table*}

\vspace{-0.6cm}
\begin{eqnarray}
\kappa(\mathbf{Y^T}(\mathbf{YY^T})^{-1}) &\leq & \kappa(\mathbf{Y})^3
\end{eqnarray}
So minimizing the condition number of SHM is equal to minimizing the condition number of the decoding matrix given in Equation \ref{SD} and \ref{MMD}. Therefore, a well conditioned SHM will avoid perturbation errors in both the encoding and decoding process. Further the coefficients of the reproduced sound field $\mathbf{b(t)}$, and the actual sound fields $\mathbf{\tilde{b}(t)}$ exhibit an error. The normalized error exhibited between them can be calculated as follows. 
\begin{equation}
\mathbf{\xi(t)}=\frac{||\mathbf{b(t)}-\mathbf{\tilde{b}(t)}||_2}{||\mathbf{\tilde{b}(t)}||_2}
\label{pwerror}
\end{equation}
The plane wave reproduction  error depends on the the dimension of SHM as given in Equation \ref{encode}. For $Q>>(N+1)^2$, reproduction error $\mathbf{\xi(t)}$ will be nearly equal to zero.  But as plane wave composes of spherical harmonic terms of infinite order, it is indeed necessary to check the error performance for higher orders. For $Q<(N+1)^2$, the normalized Euclidean error as given in Equation \ref{pwerror} would be minimum, if the SHM $(\mathbf{Y})$ exhibits orthogonal spherical harmonic vectors. A well conditioned  SHM can in fact result in a better orthogonality among spherical harmonic vectors and thereby minimize the plane wave reproduction error. Therefore in the following section, obtaining the well conditioned SHM using the method proposed in Section \ref{LAcond} would be discussed. \\


\subsubsection{Minimization of SHM Condition Number for Identifying the Loudspeaker Geometry} 
\label{condfibo}
Consider $Q$ points distributed over a sphere in a Fibonacci lattice. Azimuthal and elevation angles for this $Q$ points can be obtained as \cite{Gonz치lez2009},

\begin{equation}
\begin{aligned}
 c_1 &= \frac{\sqrt 5-1}{2} \\
 \phi_i&=(2\pi\cdot c_1 \cdot i) \mod 2\pi \quad  i=1 \cdots Q \\
 \theta_i &= \sin^{-1}\{\frac{2i}{Q}-1\} \quad i=1 \cdots Q
\end{aligned}
\end{equation}
where, $\mod$ is the modulo operation. Among this $Q$ sampling points over a sphere, $Q'$ sampling points which result in a well conditioned SHM has to be identified. These sampling points are obtained by solving the optimization problem given in Equation \ref{optfinal} with an additional regularization constraint as given below.

\begin{figure}
  \centerline{\includegraphics[width=8cm,height=6.5cm,trim=500 180 500 170,clip]{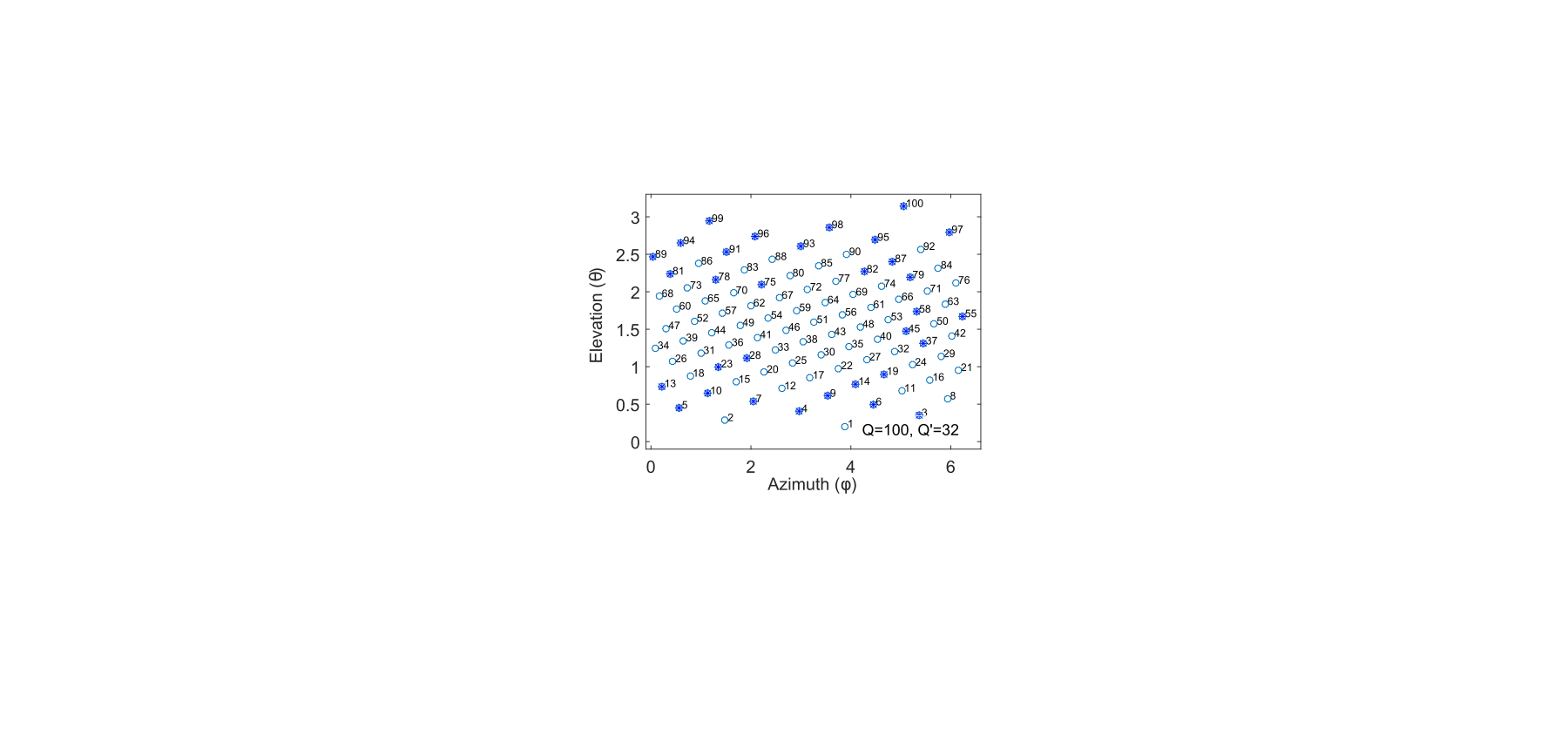}}
  \vspace{-1.3cm}
  \caption{Figure illustrating $Q'=32$ optimum spatial sampling points (filled circle) obtained using the proposed method among $Q=100$ spatial sampling points (empty circles) of the Fibonacci Lattice. }
  \label{fig:sam_points}
\end{figure}

\vspace{-0.4cm}

 \begin{equation}
\begin{aligned}
\mathbf{\Delta}^*= \quad & \underset{\mathbf{\Delta, diagonal}}{\text{minimize}}
& & \lambda_{max}(\mathbf{Y\Delta Y^{H}})\\
&\text{subject to}
& &\delta_{ii} \in {\{0,1\}} \quad  \forall i \in \{1 \cdots Q\}\\
& & &\sum_{i=1}^{Q} \delta_{ii} =Q' \\
& & &\lambda_{min}(\mathbf{Y\Delta Y^{H}}) > \eta^*
\end{aligned}
\label{optfinalSHM}
\end{equation}

\begin{equation}
\kappa^*=\sqrt{\frac{\lambda_{max}(\mathbf{Y\Delta}^* \mathbf{Y^{H}})}{\lambda_{min}(\mathbf{Y\Delta}^* \mathbf{Y^{H}})}}
\label{conddSHM}
\end{equation}
where, $\mathbf{Y}$ is the SHM of $P$ rows and $Q$ columns, $\eta^*$ in Equation \ref{optfinalSHM} can be obtained using Algorithm \ref{alg1} and Equation \ref{opti_eta}. The problem posed in Equation \ref{optfinalSHM} is first solved here for various values of $Q$. Figure \ref{fig:mineig_cond}  illustrates the minimum eigenvalue and condition number of sub matrix $(\mathbf{Y\Delta}\mathbf{Y^{H}})$ across various  lower bounds for selecting  $Q'=32$ optimal spatial points from a total of $Q=100$ sampling points. Further it can be observed from Table \ref{table1} that, number of intervals with constant condition number or the number of transition points $(R)$ is different for different values of $Q$. But as claimed in Section \ref{optLB}, for any integer value of $Q$, the value of $R$ is found to be finite and small for SHM. The length of each interval is also arbitrary and depends only on the minimum eigenvalue obtained by solving the optimization problem at each transition. It can also be observed that, condition number attains a minimum value for a particular interval of $\eta$ indicated by an arrow mark in Figure \ref{fig:mineig_cond}. The condition number values obtained for optimum lower bound $\eta^*$ for an SHM of order $N=3$ is tabulated in Table \ref{table1}. Choosing any value for $\eta$ in this interval result in a sub-matrix of SHM which results in minimum condition number. $Q'$ columns of these resultant sub-matrix correspond to a $Q'$ spatial directions.  $Q'$ optimum sampling points obtained are a subset of $Q$ sampling points of Fibonacci lattice. These optimal sampling points obtained for $Q=100,Q'=32$ are illustrated in Figure \ref{fig:sam_points}. It can be observed that optimum sampling points are away from the plane $\theta=\pi/2$ and spread uniformly along entire azimuth range.

\subsection{Spatial Sampling Configurations for HRTF Measurement using SHM Condition Number Minimization}
\label{HRTFsec}
Spherical harmonics are widely used for representation of acoustic fields over a sphere. HRTFs are considered to be samples of a valid acoustic field. So they can be also be expressed in spherical harmonics \cite{Evans,zotkin2009regularized}. These representations not only capture the HRTFs compactly in few coefficients but also enable to interpolate and extrapolate HRTFs \cite{durai_int_ext}. HRTFs of various directions can be represented in spherical harmonic functions as \cite{Avni,abhaya_insights},

\vspace{-0.3cm}

\begin{eqnarray}
& \mathbf{H(k)}=\mathbf{Y^TH_{nm}(k)}   \label{HRTFeq}\\ \nonumber
& \mathbf{H(k)}=[H(k,\theta_1,\phi_1) \hspace{0.1cm}  H(k,\theta_2,\phi_2) \hspace{0.1cm}  \cdots \hspace{0.1cm}  H(k,\theta_Q,\phi_Q)]^T_{Q \times 1} \nonumber\\ \nonumber
& \mathbf{H_{nm}(k)}=[H_{00}(k) \hspace{0.1cm}  H_{1(-1)}(k) \hspace{0.1cm} \cdots \hspace{0.1cm} H_{NN}(k)]^T_{(N+1)^2 \times 1} \nonumber\\ \nonumber 
& k=\frac{2\pi f}{c}, \quad c=340 \hspace{0.1cm} m/s \nonumber 
\end{eqnarray}
$H(k,\theta_j,\phi_j)$ represents the head related transfer function of wave number $k$ and spatial direction $(\theta_j,\phi_j)$.  $H_{nm}(k)$ represents spherical harmonic coefficients of degree $n$ and order $m$, and $\bold{Y}$ represent spherical harmonic matrix as given in Equation \ref{SHM}. The spherical harmonic coefficients $H_{nm}(k)$ can be obtained using the least square solution of Equation \ref{HRTFeq}. HRTFs of any arbitrary direction can now be obtained using the spherical harmonic expansion as follows:

\begin{figure*}[t!]
  \centerline{\includegraphics[width=18cm,height=6cm,trim=30 50 30 50,clip]{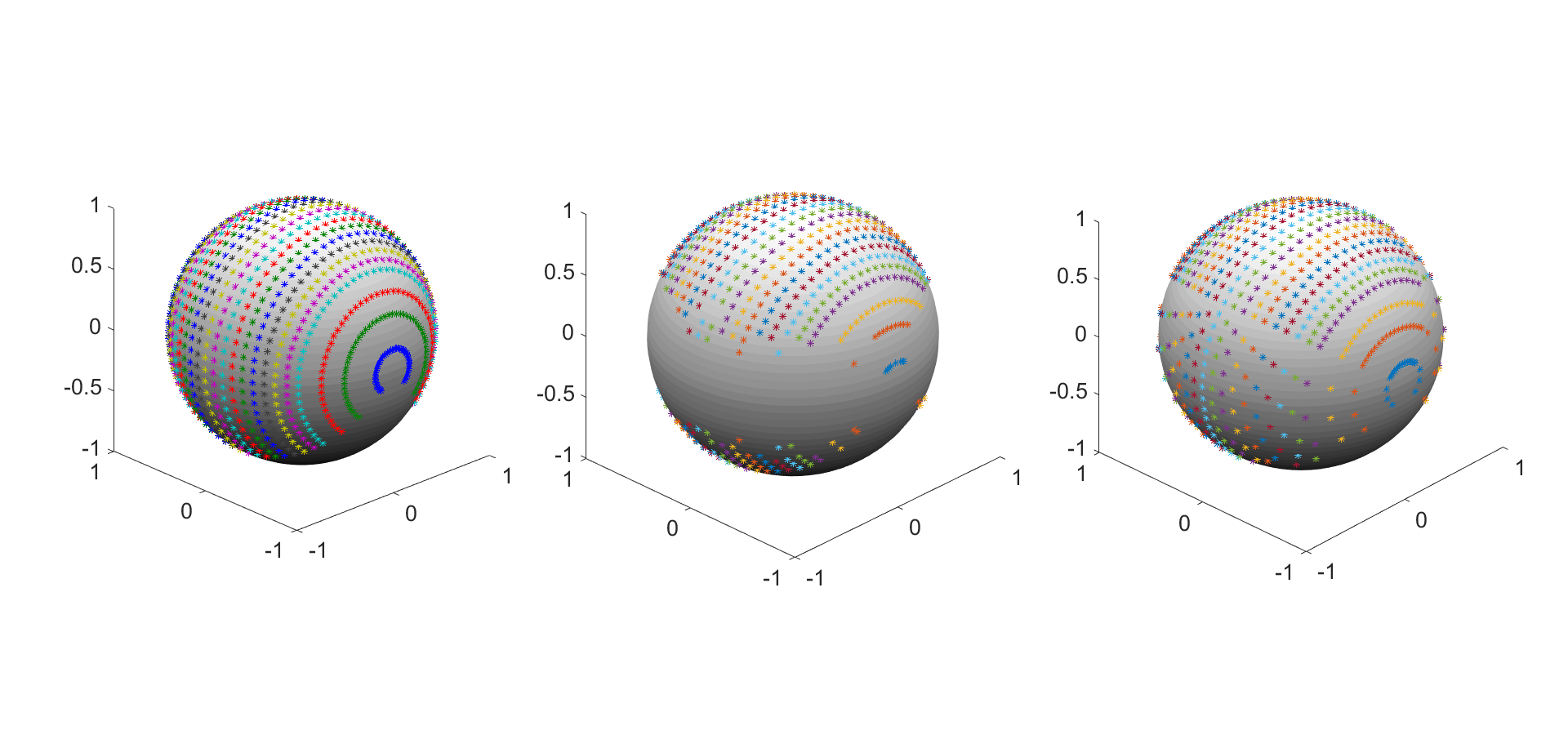}}
  \vspace{-1cm}
  \caption{Distribution of sampling points on a sphere for various methods (a)  original CIPIC data of 1250 sampling points.  (b) 640 optimum sampling points obtained using the proposed method.  (c) The modified CIPIC configuration of 850 sampling points.}
  \label{fig:cipic_allfigs}
\end{figure*}

\vspace{-0.15cm}
 
 \begin{equation}
 \tilde{H}(k,\theta,\phi)=\sum_{n=0}^{N} \sum_{m=-n}^{n} H_{nm}Y_n^m(\theta,\phi).
 \label{HRTFinterp_eq}
 \end{equation}
where,  $\tilde{H}(k,\theta,\phi)$  is the interpolated HRTF obtained for direction $(\theta,\phi)$. It can be observed from Equation \ref{HRTFeq} that,
the accuracy of the interpolated HRTF depends up on the sampling points chosen for SHM. It is important to choose a configuration for spatial sampling points which has a better condition number for SHM $(\mathbf{Y})$. Otherwise the errors incurred during measurement of HRTFs would be over sensitive during spherical harmonic transformation. But in order to reduce the complexity of HRTF measurement process,  circular hoops were earlier utilized for simultaneous measurement of HRTFs from various directions. So the optimal spatial sampling configuration  for HRTF measurement should not only reduce the condition number of the spherical harmonic matrix but also enable us to take advantage of circular hoops. However traditional sampling schemes like Interaural sampling scheme or Gaussian sampling scheme leads to an ill-conditioned SHM. For example, condition number of CIPIC configuration is 338.0216. This value can be reduced by selecting a subset of optimal spatial sampling points  discussed as follows.

Consider $Q$ spatial sampling points of the 3D space. $Q'$  sampling points $(Q'<<Q)$, resulting in a well conditioned SHM can be identified by solving the optimization problem given in Equation \ref{optfinalSHM}. However, it is sometimes necessary to constrain the total number of sampling positions in a circular hoop. For example, CIPIC configuration has dense sampling at the left and right poles. The number of sampling positions can be restricted based on the relevance of HRTFs in spatial sound perception. Thus the newly constrained optimization problem is given as,
 
\vspace{-0.2cm}
 
\begin{equation}
\begin{aligned}
\mathbf{\Delta}^*= \quad & \underset{\mathbf{\Delta, diagonal}}{\text{minimize}}
& & \lambda_{max}(\mathbf{Y\Delta Y^{H}}) \\
&\text{subject to}
& &\delta_{ii} \in {\{0,1\}} \quad  \forall i \in \{1 \cdots Q\}\\
& & &\mathbf{U}\bm{\delta} \leq \mathbf{v} \\
& & &\sum_{i=1}^{Q} \delta_{ii} =Q' \\
& & &\lambda_{min}(\mathbf{Y\Delta Y^{H}}) > \eta^* 
\end{aligned}
\label{optfinalHRTF}
\end{equation}

\usetikzlibrary{matrix,decorations.pathreplacing}

\begin{tikzpicture}
\node at (-4.2,0) {$\mathbf{U}=$};
   \matrix (m) [matrix of math nodes,left delimiter=[,right delimiter={]}] {
1 & 1 &  \cdots & 1 & 0 & 0 & \cdots & 0 & 0 & \cdots & 0 & 0 &\cdots & 0 \\ 
0 & 0 &  \cdots & 0 & 1 & 1 & \cdots & 1 & 0 & \cdots & 0 & 0 &\cdots & 0 \\ 
\cdot & \cdot &  \cdots & \cdot & \cdot & \cdot & \cdots & \cdot & \cdot & \cdots & \cdot & \cdot &\cdots & \cdot \\  
\cdot & \cdot &  \cdots & \cdot & \cdot & \cdot & \cdots & \cdot & \cdot & \cdots & \cdot & \cdot &\cdots & \cdot \\
0 & 0 &  \cdots & 0 & 0 & 0 & \cdots & 0 & 0 & \cdots & 1 & 1 &\cdots & 1 \\
    };
\end{tikzpicture}


\begin{equation}
\bm{\delta}=[\delta_{11} \quad \delta_{22} \quad \cdots \quad \delta_{QQ}]^T
\end{equation}
where $\bm{\delta}$ is a vector with diagonal elements of matrix $\mathbf{\Delta}$. $\mathbf{U}$ is a binary matrix with $J$ rows and $Q$ columns, where $Q=JK$. Here $J$ indicate the number of circular hoops and $K$ indicate number of sampling positions in each circular hoop. The $i^{th}$ element of vector $\mathbf{v}$ indicate the upper bound for number of sampling points in the $i^{th}$ circular hoop. Solving this above optimization problem  results in $Q'$ optimal sampling points which reduces the condition number and constrain the number of sampling points in each circular hoop. Simulations are performed for sampling points of CIPIC data. Among the total 1250 sampling points, with 25 circular hoops and 50 sampling points per each hoop, 640 optimal sampling points are identified. The number of sampling points in each  hoop is constrained to be less than an upper bound whose values are given below.


\begin{figure*}
  \centerline{\includegraphics[width=16.5cm,height=6.5cm,trim=50 240 40 20,clip]{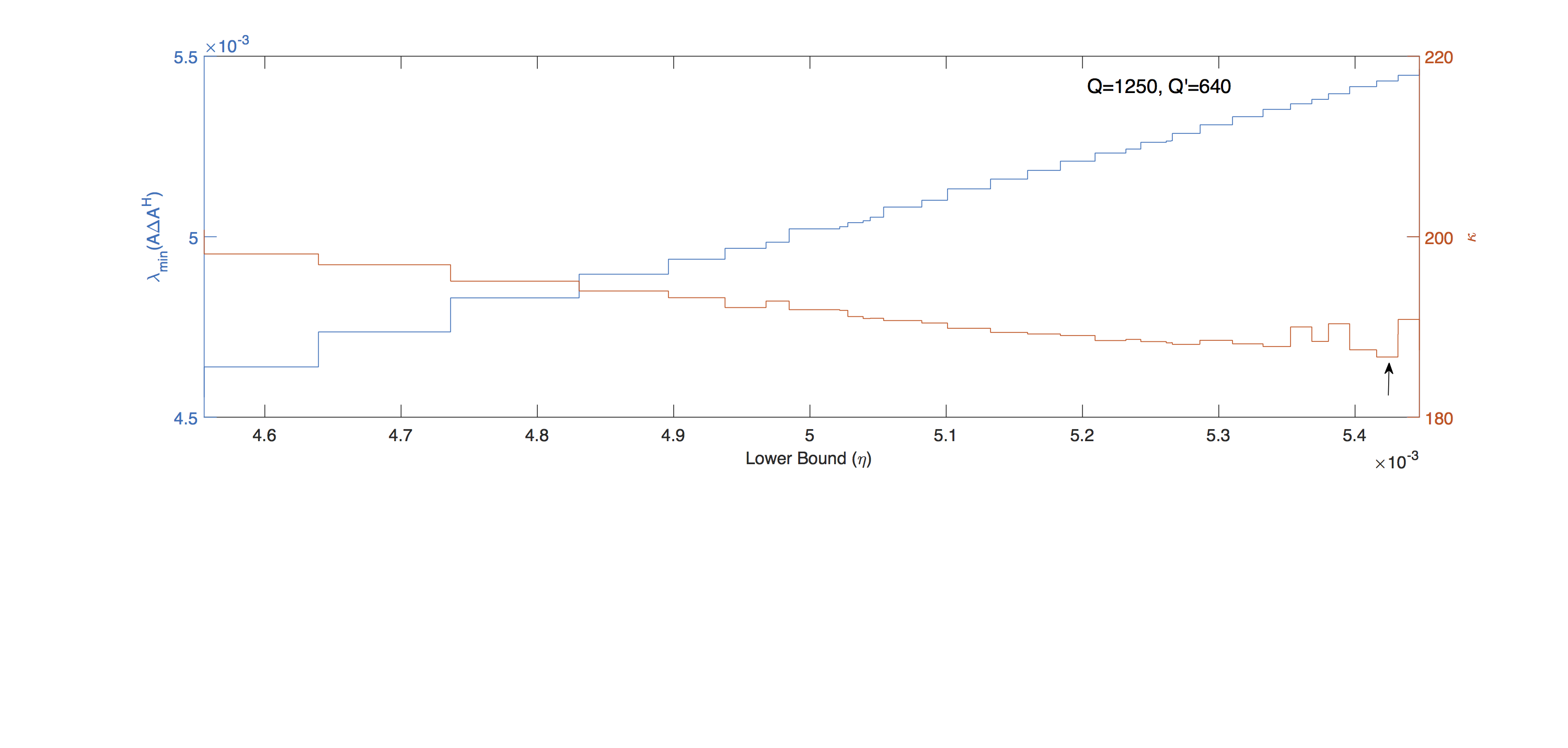}}
  \vspace{-0.35cm}
  \caption{Figure illustrating minimum eigenvalue and condition number of the sub-matrix $(\mathbf{Y\Delta}\mathbf{Y^{H}})$ for various lower bound simulated for total number of sampling points $Q=1250$ and $Q'=640$ following interaural sampling configuration of CIPIC database.}
  \label{fig:cipic_mineig_cond}
\end{figure*}
\begin{figure}
  \centerline{\includegraphics[width=9cm,height=5.5cm,trim=350 213 400 100,clip]{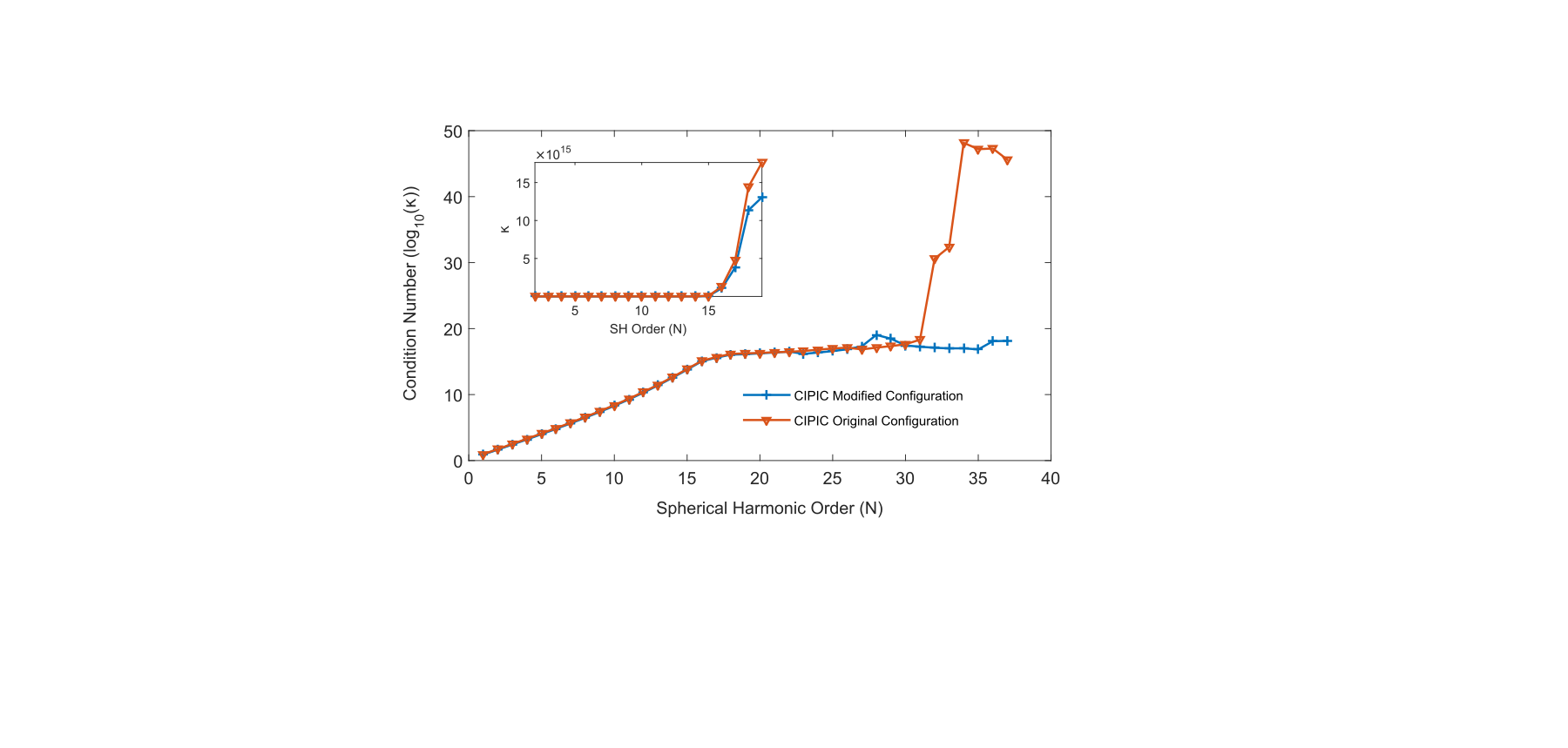}}
  \caption{Figure illustrates condition number performance of the CIPIC original configuration and CIPIC Modified configuration for various spherical harmonic orders}
  \vspace{-0.4cm}
  \label{fig:CIPICvsdegree1}
\end{figure}



\begin{eqnarray}
v_i &=& 14+3(i-1) \quad i=1 \cdots 13 \\
v_i &=& 50-3(i-13) \quad i=14 \cdots 25 
\end{eqnarray}
Figure \ref{fig:cipic_mineig_cond}, illustrates minimum eigenvalue and condition number of sub matrix $(\mathbf{Y\Delta}\mathbf{Y^{H}})$ for  $R=31$ distinct lower bounds. The condition numbers for the original sampling configuration of 1250 sampling points is found to be 338.02, and the condition number of optimum sampling configuration of 640 points is found to be 186.7. It can be observed that, there is a decrease in condition number of SHM for the latter configuration. Figure \ref{fig:cipic_allfigs} shows the original and optimal spatial sampling points for CIPIC data. It can be clearly observed that, the optimum sampling points are distributed uniformly along all the lateral angles and sparsely populated on and around the the plane $\theta=0^0$. It has to be noted that, elevation angle in the current application is measured above the $xy$ plane. Since spatial perception is an important criteria for choosing sampling points, a new configuration which distributes sampling points sparsely on and around the plane $\theta=0^0$ and distributes uniformly away from the plane is considered. The new modified configuration is shown in Figure \ref{fig:cipic_allfigs}c and its corresponding condition number is found to be 269.10. Condition number performance of the CIPIC original configuration and CIPIC Modified configuration for various spherical harmonic orders is shown in Figure \ref{fig:CIPICvsdegree1}.  Proposed method is found to be reducing the condition number of SHM, particularly for the higher spherical harmonic orders.

\section{Performance Evaluation}
\label{sec4}
Performance of the proposed method in obtaining the spatial sampling configurations for  spatial audio applications is discussed herein. Initially the practical implementation of the proposed method is presented. Further the SHM condition number performance and the plane wave reproduction error is analysed for various spherical harmonic orders. Finally HRTF interpolation accuracy is evaluated for various interaural sampling configurations and spherical harmonic orders.


\subsection{Implementation}
\label{prac_imple}
\vspace{-0.1cm}
Practical implementation of the proposed algorithm for condition number minimization  is performed using numerical optimization. The fundamental problem formulated in Equation \ref{opt2} falls in to the category of mixed integer programming with binary variables. The objective function and the constraints are all convex except that the variables are constrained to be binary. This problem was complex to solve using a traditional linear or convex programming tools. However the recent advancements in optimization, particularly in the domain of mixed integer programming resulted in a number of optimization toolboxes that are freely available in the public domain \cite{Lofberg2004}. YALMIP \cite{Lofberg2004} is one such toolbox which can be interfaced to MATLAB and also enable the user to choose appropriate solver for a particular purpose. Sedumi 1.3 \cite{sedumi1.3}  is a mixed integer programming solver that is freely available for academic purpose. This solver is used  to solve all the mixed integer optimization problems that are formulated in this work. It is important to note that, the mixed integer toolboxes need not necessarily result in a global optimum solution. Nevertheless, this work attempts at achieve the best possible reduction in condition number maintaining the application specific requirements.  It is also to be noted that, the fundamental problem posed in Equation \ref{opt2} contains strict inequalities $(>)$ on the lower bound of the minimum eigenvalue. But in general, the optimization solvers can only solve non-strict inequalities $(>=)$. Hence a small value of $\epsilon=10^{-7}$ is used to convert strict inequalities in to non strict inequalities as shown below.


\begin{eqnarray}
\text{Strict Inequality:} & \lambda_{min}(\mathbf{A\Delta A^{H}}) > \eta  \\
\text{Non-strict Inequality:} & \lambda_{min}(\mathbf{A\Delta A^{H}}) >= \eta+\epsilon
\end{eqnarray}
The simulations of the proposed algorithm is performed using MATLAB 2015 on a 64 bit PC of 8GB RAM. Most of the performance analysis is performed for a maximum $Q$ value of $2000$. It has been found that, the optimization problem for $Q>2000$ and $N>5$ results in insufficient memory issues for the above configurations. High end computers can be utilised to extract the results of the proposed algorithms for higher number of sampling points and spherical harmonic orders.

\begin{table*}[b!]
\centering
\caption{Table listing the condition number and D-measure values for proposed and conventional methods for various spherical harmonic orders $N$, and sampling points $Q'$ selected from a total of $Q=100$ sampling points.}

\vspace{0.3cm}
\label{tablefinal}
\begin{tabular}{cccccccccccc}
 \hline \hline 
\multicolumn{2}{c}{}                            & \begin{tabular}[c]{@{}c@{}}$Q'=6$\\ $N=1$\end{tabular} & \begin{tabular}[c]{@{}c@{}}$Q'=8$\\ $N=1$\end{tabular} & \begin{tabular}[c]{@{}c@{}}$Q'=12$\\ $N=2$\end{tabular} & \begin{tabular}[c]{@{}c@{}}$Q'=18$\\ $N=2$\end{tabular} & \begin{tabular}[c]{@{}c@{}}$Q'=20$\\ $N=3$\end{tabular} & \begin{tabular}[c]{@{}c@{}}$Q'=24$\\ $N=3$\end{tabular} & \begin{tabular}[c]{@{}c@{}}$Q'=30$\\ $N=4$\end{tabular} & \begin{tabular}[c]{@{}c@{}}$Q'=32$\\ $N=4$\end{tabular} & \begin{tabular}[c]{@{}c@{}}$Q'=36$\\ $N=4$\end{tabular} & \begin{tabular}[c]{@{}c@{}}$Q'=50$\\ $N=4$\end{tabular} \\ \hline \\
\multirow{2}{*}{Proposed}  & $log_{10}(\kappa)$ & 0.422                                             & 0.444                                             & 1.478                                              & 1.269                                              & 2.560                                              & 2.331                                              & 4.033                                              & 3.647                                              & 3.455                                              & 3.427                                              \\  
                           & $D$                & 0.080                                             & 0.088                                             & 1.046                                              & 1.199                                              & 0.352                                              & 0.981                                              & 0.723                                              & 0.968                                              & 1.070                                              & 0.683                                              \\ \hline \\
\multirow{2}{*}{T-design}  & $log_{10}(\kappa)$ & 0                                                  & 0                                                  & 0                                                   & -                                                   & -                                                   & 0                                                   & -                                                   & -                                                   & 0                                                   & -                                                   \\  
                           & $D$                & 0                                                  & 0                                                  & 0                                                   & -                                                   & -                                                   & 0                                                   & -                                                   & -                                                   & 0.002                                               & -                                                   \\ \hline \\
\multirow{2}{*}{Fibonacci} & $log_{10}(\kappa)$ & 1.227                                             & 0.828                                             & 2.256                                              & 1.935                                              & 4.061                                              & 3.424                                              & 5.172                                              & 5.581                                              & 4.575                                              & 3.885                                              \\  
                           & $D$                & 0.095                                             & 0.077                                             & 0.048                                              & 0.032                                              & 0.029                                              & 0.024                                              & 0.019                                              & 0.018                                              & 0.016                                              & 0.012                                              \\ \hline \\
\multirow{2}{*}{Gaussian}   & $log_{10}(\kappa)$ & -                                                  & 0.383                                             & -                                                   & 16.037                                              & -                                                   & -                                                   & -                                                   & 16.752                                              & -                                                   & 16.496                                              \\  
                           & $D$                & -                                                  & 0.036                                             & -                                                   & 0.065                                              & -                                                   & -                                                   & -                                                   & 0.176                                              & -                                                   & 0.292                                             \\ \hline \hline
\end{tabular}
\end{table*}

\subsection{Computational Complexity Analysis}
The importance of the proposed method is mainly due its capability to reach an optimal solution in a practically acceptable amount of time for $Q<2000, N<5$. The exhaustive search for finding an optimal solution is practically impossible. For example, a matrix with $Q=100$, $N=3$ and $Q'=32$ would require $10^{20}$ hours for computing the optimum sub-matrix and minimum condition number. Where as, the proposed method can reach to an optimal solution in less than 12 hours. In fact, the proposed method would consume less time for SHM than any other matrix, because of the properties of spherical harmonic functions as discussed below.

\vspace{-0.4cm}

\begin{eqnarray}
Y_n^m(-\theta,\phi)&=&Y_n^m(\theta,\phi)\\
Y_n^m\Big(\frac{\pi}{2}+\theta,\phi \Big)&=&Y_n^m\Big(\frac{3\pi}{2}-\theta,\phi \Big)
\end{eqnarray}

\vspace{-0.2cm}
These properties of spherical harmonics result in similar columns for SHM. In general  concatenating any number of column vectors which are identical to the already existing column vectors of a matrix does not alter the condition number. So computational complexity for identifying a sub-matrix which result in minimum condition number can be reduced for SHM using the aforementioned properties. For example, consider CIPIC configuration of 1250 sampling points on a sphere. Number of sampling points which result in similar spherical harmonic coefficients is 909. So in this case, the computational complexity is reduced by one-fourth of the actual time consumed.

Computational complexity of the proposed method also depends on the upper bound of the minimum eigen value. A simple function depending on the trace of the matrix $\mathbf{YY^H}$ is considered as an upper bound for $\eta$ in this work. The values of $\eta$ are increased stepwise from lower bound $0$ to upper bound $\frac{Tr(\mathbf{YY^H})}{P}$ as given in Equation \ref{stepeta}. However the algorithm proposed in this work can also terminate without actually reaching the upper bound. This can happen if the problem formulated in Equation \ref{optdiscrete} turns in to an infeasible optimization problem for a particular value of $\eta$. The number of transition points $R$ depends on the number of steps the proposed algorithm has taken to either reach an infeasible state or the upper bound $\frac{Tr(\mathbf{YY^H})}{P}$. Time complexity of the proposed algorithm depends on the value of $R$ and the time taken for the optimization problem to reach  an optimal solution. So tighter upper bound for $\eta$ would reduce the number of transition points $R$ and thereby reduce the time complexity of the proposed algorithm.



\begin{figure*}[t]
  \centerline{\includegraphics[width=16cm,height=5cm,trim=0 200 0 140,clip]{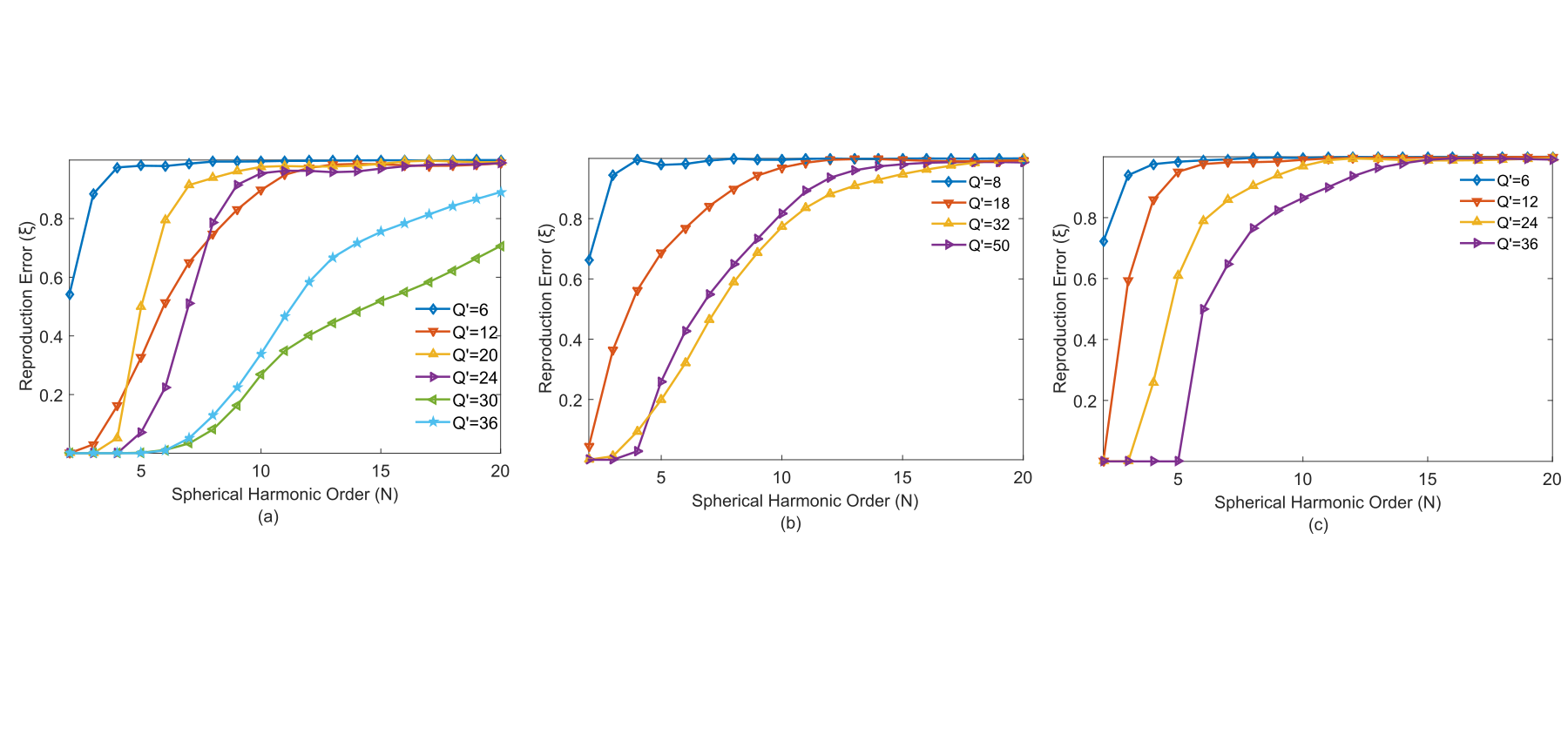}}
  \vspace{-0.3cm}
   \caption{Figure illustrating normalised  reproduction error $(\xi)$ between sound field generated by a plane wave source located at $(\tilde{\theta_1},\tilde{\phi_1})=(\pi/4,\pi/4)$ and sound field reproduced by loudspeakers for various spherical harmonic order $(N)$ and sampling points $Q'$ arranged in (a) Proposed configuration, (b) Gaussian configuration (c) T-designs.}
  \label{fig:RError}
\end{figure*}

\subsection{Condition Number Analysis of Various Sampling Schemes for Spatial Sound Reproduction}

 The performance of various sampling configurations are evaluated in terms of conditioning behaviour of SHM and uniformity of spatial sampling points. D-measure as discussed in section \ref{uni_D_measure} is used to quantify the uniformness of spatial sampling points. Table \ref{tablefinal} depicts the condition number and D-measure values for various number of sampling points $(Q')$ and spherical harmonic orders $(N)$ across three different configurations namely, T-design, Fibonacci, Gaussian and the proposed configuration.  It can be observed that T-designs  result in the lowest condition number and D-measure values if there exists a configuration for the particular value of number of sampling points $(Q')$ and spherical harmonic orders $(N)$. For example, it can be noted from the Table \ref{tablefinal} that, for some of $(Q',N)$ values, T-design configuration does not exist. Fibonacci lattice as discussed in Section \ref{condfibo} provides nearly uniform distribution for sampling points. This can also be noted from D-measure values of Fibonacci lattice in Table \ref{tablefinal}. But it can be observed that, the condition number values given in Table \ref{tablefinal} in logarithmic scale are found to be higher as compared to the proposed method. Gaussian distribution of sampling points are equisampled in azimuthal angles, and exhibits lesser condition number and D-measure values for $Q'=8, N=1$ as compared to the proposed method. But for $Q'>8$, $N>1$ both the condition number and D-measure values of Gaussian configuration are found to be higher which can be noted from Table \ref{tablefinal}. The sampling points obtained from the proposed configuration exhibits lesser condition number than the Fibonacci and Gaussian distributions. This result comes at a cost of increase in non-uniformity as compared to Fibonacci configuration. It can be understood from the above mentioned observations that the proposed method is able to result in a well conditioned SHM as compared to Fibonacci and Gaussian configurations. Unlike T-design, proposed method is able to generate sampling points for any number of sampling positions in the 3D space.

\subsection{Plane-wave Reproduction Error Analysis of Various Loudspeaker Geometries}
\label{PW_repro}


In section \ref{SHM_pw_LS}, the error exhibited between the actual sound field and the reproduced sound field using $Q$ loudspeakers is given in Equation \ref{pwerror}. Sound field generated by a plane-wave source from the direction $(\tilde{\theta_1},\tilde{\phi_1})=(\pi/4,\pi/4)$ is synthesized using Equation \ref{repro_pw}. This sound field is reproduced by the loudspeakers arranged in various geometries. The normalised reproduction error $(\xi)$ obtained for various loudspeaker geometries like T-designs, Gaussian and the Proposed sampling schemes are considered for evaluations. Sampling positions of T-designs obtained for $Q'=\{6, 12, 24, 36\}$ are used for computing the plane wave reproduction error $\mathbf{\xi}$. As total number of sampling points of a Gaussian grid on sphere is multiple of $2(N+1)^2$, error $\mathbf{\xi}$ is computed for $Q'=\{8,18,32,50\}$. For the proposed method, error analysis is performed for sampling points obtained for $Q'=\{6,12,20,24,30,36\}$. For each of these sampling schemes, error $\mathbf{\xi}$ is computed across various spherical harmonic orders $(N)$. The normalised error obtained for all the three configurations is illustrated in Figure \ref{fig:RError}. It can be observed that, for $Q' \geq (N+1)^2$, T-designs and the proposed method exhibit nearly zero error. For $Q'<(N+1)^2$, the error increases with the spherical harmonic order. But the rate of increase in error is found to be less for the proposed method. This can be noted from Figure \ref{fig:RError}a for $Q'=30$, and $Q'=36$. Irrespective of the choice of sampling scheme, plane wave reproduction using limited number of sampling points exhibit a maximum error for higher spherical harmonic orders. In order to study the dependence of plane wave source direction on the reproduction error, simulations are performed for 648 (36 azimuthal and 18 elevation)  different source directions independently. Number of directions (in percentage) proposed and T-designs sampling scheme exhibiting lesser reproduction error for different regularization parameter values  is given in Table \ref{table_percent}. It can be noted that, across all possible values of $(Q',N)$ proposed method has lesser error for more number of directions.

%

\begin{table}[h!]
\centering
\caption{Number of directions (in \%) exhibiting lesser reproduction error by \\ proposed and T-design and T-design sampling schemes}

\vspace{0.2cm}
\label{table_percent}
\begin{tabular}{ccccc}
\hline \hline \\
                                                       & $Q'$=6,N=2 & $Q'$=12, N=3 & $Q'$=24, N=4 & \multicolumn{1}{c}{$Q'$=36, N=6} \\ \hline \\
T-design                                               & 46.44\   & 27.62\       & 36.57\    & 47.84\                        \\ \hline  \\
\begin{tabular}[c]{@{}c@{}}Proposed \end{tabular}                                             & 53.56     & 72.38          & 63.43      & 52.16                           \\ \hline \\
\end{tabular}
\end{table}

\vspace{-0.5cm}
\subsection{Performance Analysis of HRTF Interpolation for Various Interaural Sampling Configurations}
In section \ref{HRTFsec}, it has been noted that, choice of different sampling configurations will alter the conditioning behaviour of SHM and in turn have an effect on the accuracy of the interpolated HRTFs. Further, it has also been observed that, in the interaural sampling configurations, choosing sparse sampling at the left and right poles results in a better condition number for SHM. Considering this observations, the error incurred in the interpolated HRTFs for two different configurations will be discussed herein. The first configuration is the Equi-sampled CIPIC Configuration (ECC) with elevation angles spaced by $11.25^0$ as compared to the original CIPIC angular spacing of $5.625^0$ \cite{cipic}.  This generates $625$ samples with $25$ samples for each interaural circle. The second configuration is the Modified CIPIC Configuration (MCC) generated by choosing sparse sampling towards the left and right poles as discussed in Figure \ref{fig:cipic_allfigs}. In total $625$ number of samples are chosen with a variable number of samples for each of $25$ interaural circles. The elevation angular separation for each of the interaural circle in the range between $-45^0$ to $45^0$ and $135^0$ to $230.625^0$ is given in Table \ref{tab:cipic}. 

\begin{table}[h!]
\centering
\caption{Table enumerates elevation angular spacing followed in MCC}
\vspace{0.3cm}
\label{tab:cipic}
\begin{tabular}{|c|c|c|c|c|c|}
\hline
\begin{tabular}[c]{@{}c@{}}Lateral\\ angles\end{tabular}               & \begin{tabular}[c]{@{}c@{}}$\pm80^0$\\ $\pm65^0$\\ $\pm55^0$\end{tabular} & \begin{tabular}[c]{@{}c@{}}$\pm45^0$\\ $\pm40^0$\\ $\pm35^0$\\ $\pm30^0$\end{tabular} & \begin{tabular}[c]{@{}c@{}}$\pm25^0$\\ $\pm20^0$\\ $\pm15^0$\end{tabular} & \begin{tabular}[c]{@{}c@{}}$\pm10^0$\\ $\pm5^0$\end{tabular} & $0^0$     \\ \hline
\begin{tabular}[c]{@{}c@{}}Elevation \\ angular\\ spacing\end{tabular} & $33.75^0$                                                                     & $28.125                                                                               ^0$ & $22.5^0$                                                                      & $16.875 ^0$                                                     & $11.25^0$ \\ \hline
\end{tabular}
\end{table}

\begin{figure*}[t!]
\hspace{0.5cm}
\begin{minipage}[h]{0.32\linewidth}
  \centering
  \centerline{\includegraphics[width=6cm,height=4.5cm]{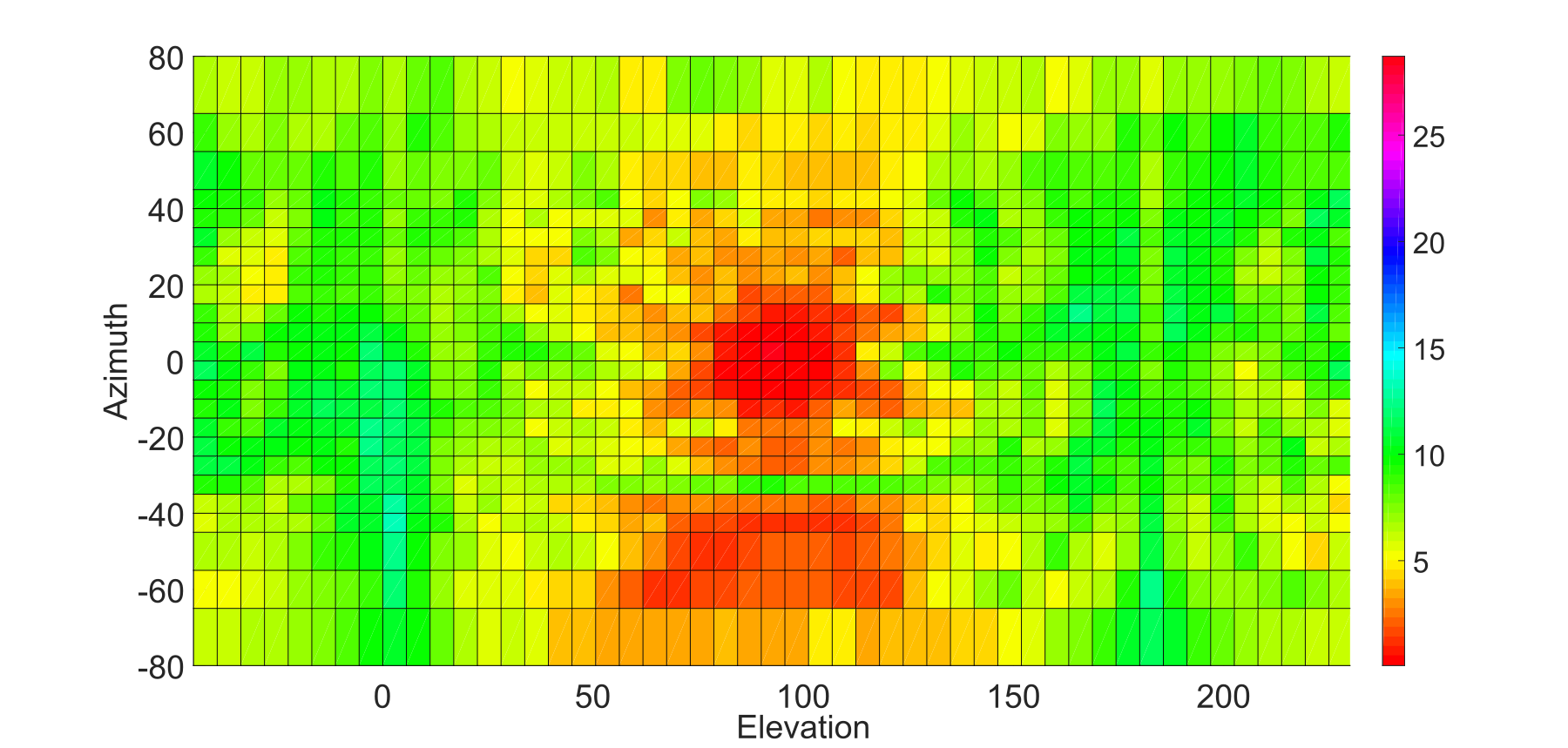}}
\centerline{$(a)$}\medskip
\end{minipage}
\begin{minipage}[h]{0.32\linewidth}
  \centering
  \centerline{\includegraphics[width=6cm,height=4.5cm]{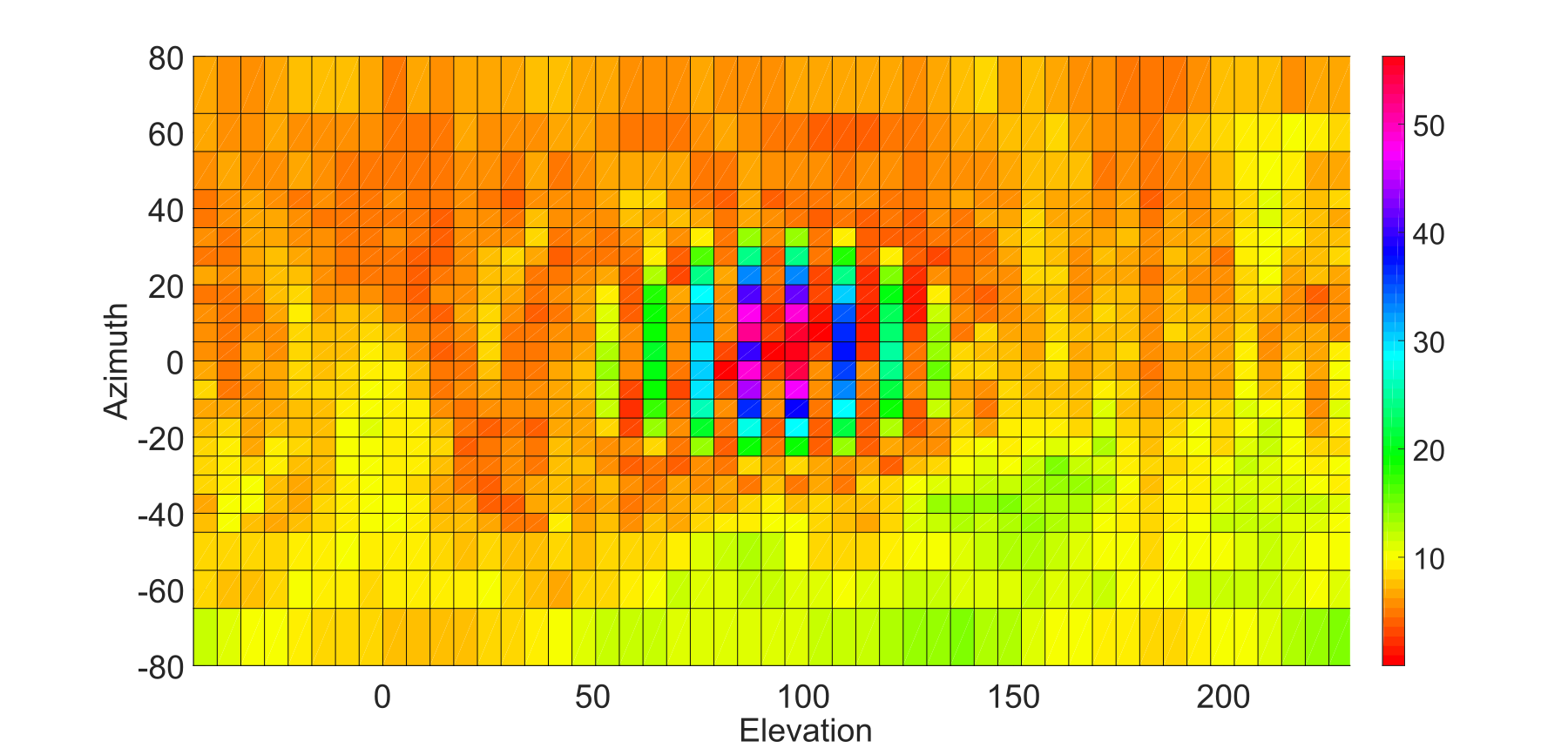}}
\centerline{$(b)$}\medskip
\end{minipage}
\begin{minipage}[h]{0.32\linewidth}
  \centering
  \centerline{\includegraphics[width=6cm,height=4.5cm]{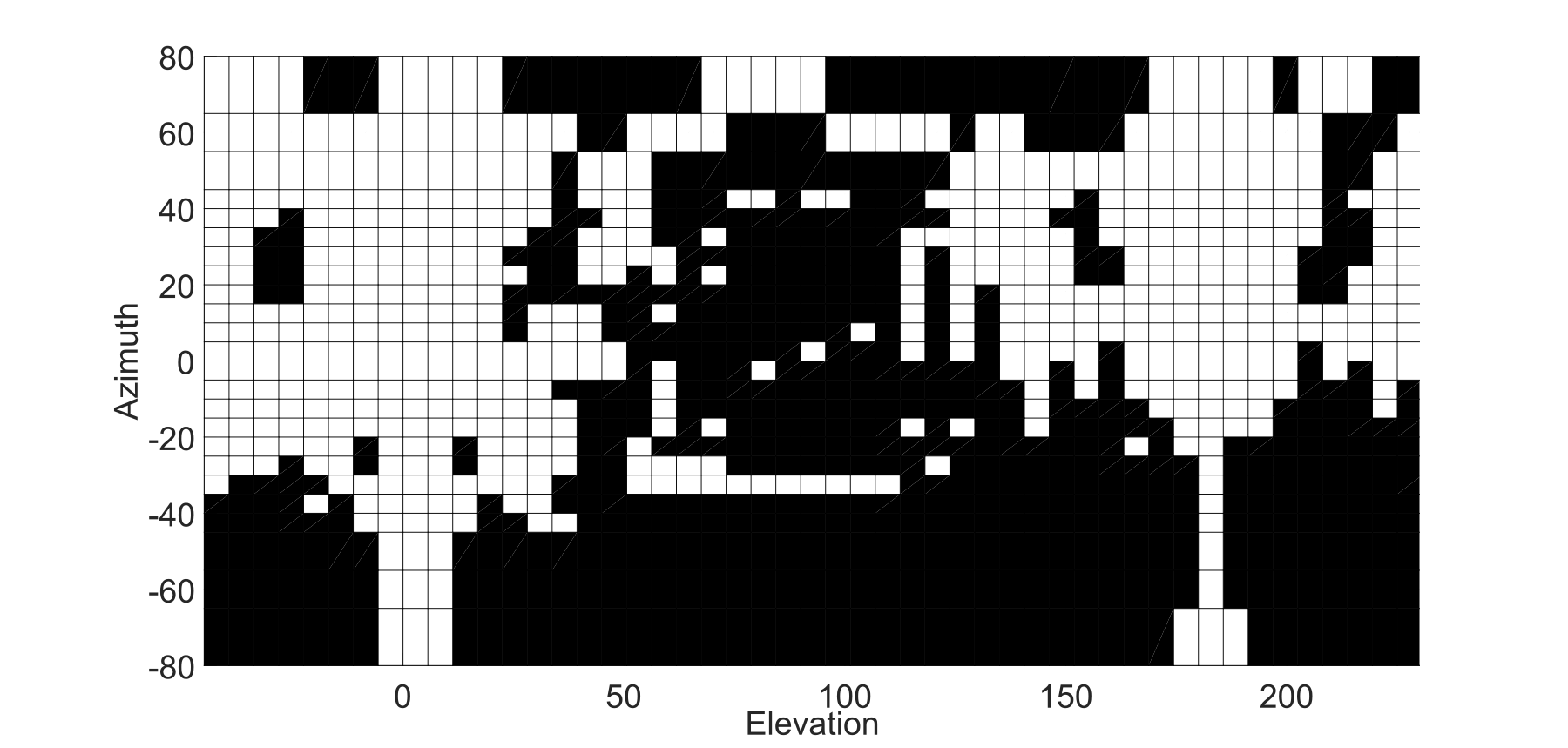}}
  \centerline{$(c)$}\medskip
\end{minipage}
\vspace{-0.4cm}
\caption{Figure illustrates LSD between the ground truth HRTFs and the interpolated HRTFs obtained for (a) Modified CIPIC Configuration (MCC), (b) Equi-sampled CIPIC Configuration (ECC) for Left ear of Subject 003 of CIPIC database. Figure (c) illustrates the regions (Black) where MCC has smaller error as compared to ECC.}
\label{fig:LSD_HRTFinterp}
\end{figure*}

Spherical harmonic coefficients $H_{nm}(k)$ are obtained for both the above configurations using the least square solution of Equation \ref{HRTFeq}. Using these coefficients HRTFs can be either reconstructed or interpolated for measured or non-measured directions respectively.  Interpolated or reconstructed HRTFs are obtained using Equation \ref{HRTFinterp_eq}. Spherical harmonic order of $N=10$ is considered for computing interpolated HRTFs.  In order to analyse the performance of interpolated HRTFs for all the angles, Log Spectral Distortion (LSD) between the ground truth HRTFs and HRTFs obtained using each of the above two methods is performed. Figure \ref{fig:LSD_HRTFinterp} illustrates the LSD computed for all directions described in the CIPIC database. It is to noted that the color bars of Figure \ref{fig:LSD_HRTFinterp} (a,b) are of different scale. It can be clearly observed that, MCC has lesser error for elevations $50^0$ to $130^0$. Further for all of the ipsilateral angles $(-80^0,0^0)$, where the HRTFs magnitude is high, MCC performs better than ECC. This can be noted from Figure \ref{fig:LSD_HRTFinterp}c, where the regions in black indicate MCC has lower error than ECC. So using this analysis it can be concluded that, a condition number improvement in SHM can result in better interpolated HRTFs.

\vspace{-0.3cm}

\section{Conclusions and Future Work}

A novel method for reducing the condition number of spherical harmonic matrix is presented in this work. Two applications of his method in the domain of spatial audio are also presented. The proposed method is able to identify the loudspeaker geometry by maintaining a trade off between reproduction accuracy and well-conditioning behaviour of SHM. The proposed method is also found to be useful in the optimal selection of spatial points  for better spherical harmonic representation of HRTFs. The primary solution proposed in this work can be further improved by identifying tighter upper bounds for minimum eigenvalue. The conditioning behaviour of SHM in the applications of spherical array processing like Beamforming and Localization will be studied as future work.


\section{Appendices}

\vspace{-0.3cm}

\subsection{{proof for  equivalence of optimization problems}}
\label{append1}


The optimization problems formulated in Equation \ref{opt1} and Equation \ref{opt2} results in same solution. If $i^{th}$ non-zero element lies at the $q^{th}$ diagonal position of $\mathbf{\Delta^*}$, then the $q^{th}$ row and $i^{th}$ column of $\mathbf{\Gamma^*}$ will contain a one non-zero element of value 1. In this manner, the elements of $\mathbf{\Delta^*}$ and $\mathbf{\Gamma^*}$ has a one-to-one correspondence. In order to prove minimization of $\frac{\lambda_{max}(\mathbf{A\Delta A^{H}})}{\lambda_{min}(\mathbf{A\Delta A^{H}})}$ and minimization of $\lambda_{max}(\mathbf{A\Delta A^{H}})$ under $\lambda_{min}(\mathbf{A\Delta A^{H}})>\eta^*$ result in same solution, consider the following line diagram.
 
 
 \begin{figure}[h]
\centering

\newcommand{\midarrow}{\tikz \draw[-triangle 90](0,0) -- +(.1,0);}
\begin{tikzpicture}[node distance=2.4cm]
\tikzstyle{arrow} = [thick,->,>=stealth]
\tikzstyle{arrow1} = [thick,<-,>=stealth]
\tikzstyle{arrow2} = [thick,-,>=stealth]
\tikzstyle{arrow3} = [dashed,-,>=stealth]

\node at (0.6,0) {}; 
   \draw (-3,-0.2)--(-3,0.2);
 \draw (-2,-0.2)--(-2,0.2);
\draw (-0.5,-0.2)--(-0.5,0.2);   
\draw (0.5,-0.2)--(0.5,0.2); 
\draw (1.5,-0.2)--(1.5,0.2); 
\draw (4,-0.2)--(4,0.2); 
 \node at (-3,0.4) {$0$};
  \node at (-2,0.4) {$\eta^*$};
  \node at (-0.5,0.4) {$\tilde{\lambda}_{min}$};
 \node at (0.5,0.4) {$\lambda_{min}^*$};
 \node at (1.5,0.4) {$\hat{\lambda}_{min}$};
  \node at (3.6,0.4) {$\frac{Tr(\mathbf{AA^H})}{P}$};
  
  \node at (-2.8,-0.3) {$\rightarrow$};
   \node at (-2.8,-0.45) {$\lambda_{min}(\cdot)$};
  \draw  (-3,0)--(3.5,0);
  \draw [arrow3] (3.6,0)--(3.8,0);
   \draw  (3.8,0)--(4,0);
\end{tikzpicture}
\end{figure}


Case 1: Consider $\eta^*=\lambda_{min}^*$. We have, $\hat{\lambda}_{min} \geq \lambda_{min}^*$.   If $\quad \hat{\lambda}_{max}<\lambda_{max}^*$ implies $\frac{\hat{\lambda}_{max}}{\hat{\lambda}_{min}}<\frac{{\lambda}^*_{max}}{{\lambda}^*_{min}}$ which is an invalid statement as $\frac{{\lambda}^*_{max}}{\lambda^*_{min}}$ should result in minimum value. Hence in the range $\lambda_{min}(\mathbf{A\Delta A^{H}}) \geq \lambda_{min}^*$, minimum value of  $\lambda_{max}(\mathbf{A\Delta A^{H}})$ is  $\lambda_{max}^*$, and also minimum value of the $\frac{\lambda_{max}(\mathbf{A\Delta A^{H}})}{\lambda_{min}(\mathbf{A\Delta A^{H}})}$ is $\frac{\lambda^*_{max}}{\lambda^*_{min}}$. Both these are obtained at the same value of $\mathbf{\Delta=\Delta^*}$. Hence optimization problems of Equation \ref{opt1} and Equation \ref{opt2} result in same solution.

 Consider $\tilde{\lambda}_{min}=\lambda_{min}(\mathbf{A \tilde{\Delta} A^{H}})$ lying in the interval $(\eta^*, \lambda^*_{min})$ as shown in the above line diagram. Let the corresponding maximum eigenvalue be $\tilde{\lambda}_{max}$. We have, $\tilde{\lambda}_{min} < \lambda_{min}^*$. Now consider two sub-cases as follows.

Case 2.1: If $\tilde{\lambda}_{max} > \lambda_{max}^*$ for every value of $\tilde{\lambda}_{min}$ lying between $[0,\lambda_{min}^*)$, then minimization of $\lambda_{max}(\mathbf{A\Delta A^{H}})$ function under $\lambda_{min}(\mathbf{A\Delta A^{H}})>\eta^*$ will result in $\lambda_{max}^*$ as the minimum value.

Case 2.2: If $\tilde{\lambda}_{max} < \lambda_{max}^*$ for atleast one $\tilde{\lambda}_{min}$ lying between $[0,\lambda_{min}^*)$, then  minimization of both the problems may not result in same solution. Because there is a possibility that, $\lambda_{max}(\mathbf{A\Delta A^{H}})$ is minimum but not the ratio $\frac{\lambda_{max}(\mathbf{A\Delta A^{H}})}{\lambda_{min}(\mathbf{A\Delta A^{H}})}$. 

So a matrix can exhibit a property discussed in  Case 2.1. If not, it will definitely exhibit the property discussed in Case 1 which indicates, indeed there is an optimal lower bound $\eta^*$ for which minimization of the $\lambda_{max}(\mathbf{A\Delta A^{H}})$ function under $\lambda_{min}(\mathbf{A\Delta A^{H}})>\eta^*$ will be equal to the minimization of the ratio  $\frac{\lambda_{max}(\mathbf{A\Delta A^{H}})}{\lambda_{min}(\mathbf{A\Delta A^{H}})}$.


\subsection{Derivation for Upper Bound $(\eta_{ub})$}
\label{append2}
Consider the sub-matrix $\mathbf{A\Delta A^{H}}$. The order of the matrix is $P \times P$. A inequality between minimum eigenvalue and  trace of the sub-matrix is given as, 

\vspace{-0.4cm}

\begin{eqnarray}
& Tr (\mathbf{A\Delta A^{H}})  \geq \lambda_{min}+\lambda_{min} \cdots \lambda_{min} =P \lambda_{min}  \nonumber \\ \nonumber 
& \lambda_{min} \leq \frac{1}{P} Tr (\mathbf{A\Delta A^{H}}) \nonumber
\label{LBUB2}
\end{eqnarray}
where $Tr(\cdot)$ denotes the trace of the matrix. However the matrix $\mathbf{\Delta}$ is unknown, therefore an upper bound which can be computed by the elements of the matrix $\mathbf{A}$ is determined. Consider $a_{ij}$, $\hat{a}_{ij}$, and $\dot{a}_{ij}$ are the elements of the $i^{th}$ row and $j^{th}$ column of matrices $\mathbf{A}$, $\mathbf{A \Delta A^{H}}$, and $\mathbf{AA^{H}}$ respectively. We have the following relations between these elements.
\vspace{-0.1cm}
\begin{eqnarray}
& \hat{a}_{ii}=\sum_{i=1}^{Q'} a_{ii}^2 <  \sum_{i=1}^{Q} a_{ii}^2=\dot{a}_{ii} \nonumber \\
& \frac{Tr (\mathbf{A\Delta A^{H}})}{P}=\frac{1}{P}\sum_{i=1}^{P} \hat{a}_{ii} \leq \frac{1}{P}\sum_{i=1}^{P} \dot{a}_{ii} = \frac{Tr (\mathbf{AA^{H}})}{P} \nonumber
\end{eqnarray}
Thus the new upper bound on the minimum eigenvalue is,
\vspace{-0.2cm}

\begin{equation}
\lambda_{min}(\mathbf{A\Delta A^{H}}) < \frac{1}{P} Tr (\mathbf{AA^{H}}) \nonumber \implies \eta_{ub}=\frac{1}{P} Tr (\mathbf{AA^{H}}).
\label{optub}
\end{equation}

\subsection{Example based illustration of Proposed Method}
\label{append3}



Consider a rectangular grid as shown below. Each column of this grid constitute possible values for $\lambda_{max}(\mathbf{A\Delta A^H})$ for a particular value of $\lambda_{min}(\mathbf{A\Delta A^H})$. Below the grid is a table listing the values of lower bound $\eta^i$, minimum and maximum eigenvalues of sub-matrix $(\mathbf{A\Delta^i A^H})$, and the corresponding condition number $\kappa^i$ for all four transitions.  Starting from a least value of lower bound, say in this example $\eta^1=0.1$, identifying the minimum value for $\lambda_{max}$ results in  $\lambda^1_{max}=1$. The corresponding  $\lambda^1_{min}$ value is $0.4$. The crucial step of the proposed method is as follows. It is unnecessary to search a minimum value for $\lambda_{max}$ for the $\eta$ lying in between $(0.1,0.4)$. It is so because, until $\eta>\lambda^1_{min}$ the minimum value for $\lambda_{max}$ is same. For instance, consider $\hat{\eta}=0.2$, minimum value of $\lambda_{max}$ for $\lambda_{min}(\cdot) > 0.1$ and $\lambda_{min}(\cdot) > 0.2$ is same, which is equal to $\lambda_{max}=1$. Therefore at $2^{nd}$ transition, until $\eta$ assumes a value greater than $\lambda_{min}^{1}=0.4$, the sub-matrix and the corresponding condition number resulted from the optimization problem for $\hat{\eta}=0.2$ would be same as that resulted for $\eta^1=0.1$. Hence the proposed algorithm will search only in the regions where there is a transition in the value of condition number. Once $\eta$ reaches its upper bound which is $\eta_{ub}=\frac{1}{P} Tr (\mathbf{AA^{H}})$, the problem become infeasible. So in the proposed algorithm the number of times the optimization problem to be solved depends on the step size for $\eta$ at each transition. As the step size increases, $\eta$ reaches its upper bound quickly which reduces the computation time. In the current example, after all the four transitions $(\lambda_{max}, \lambda_{min})=(2,0.9)$ results in the minimum condition number of $\kappa=2.2$.

\begin{figure}[h]
\centering
\newcommand{\midarrow}{\tikz \draw[-triangle 90](0,0) -- +(.1,0);}
\begin{tikzpicture}[node distance=2.4cm]
\tikzstyle{arrow} = [thick,->,>=stealth]
\tikzstyle{arrow1} = [thick,<-,>=stealth]
\tikzstyle{arrow2} = [thick,-,>=stealth]
\tikzstyle{arrow3} = [dashed,-,>=stealth]
 
 \node at (1.45,0){$\lambda_{min}(\mathbf{A\Delta A^H})$:};
 
 \node at (1.45,1.5){$\lambda_{max}(\mathbf{A\Delta A^H})$};

\node at (3.3,0.9){2};
\node at (3.9,0.9){4};
\node at (4.5,0.9){3};
\node at (5.1,0.9){1};
\node at (5.7,0.9){3};
\node at (6.3,0.9){11};
\node at (6.9,0.9){4};
\node at (7.5,0.9){2};
\node at (8.1,0.9){7};

\node at (3.3,1.5){11};
\node at (3.9,1.5){6};
\node at (4.5,1.5){6};
\node at (5.1,1.5){5};
\node at (5.7,1.5){4};
\node at (6.3,1.5){9};
\node at (6.9,1.5){10};
\node at (7.5,1.5){4};
\node at (8.1,1.5){6};

\node at (3.3,2.1){8};
\node at (3.9,2.1){9};
\node at (4.5,2.1){8};
\node at (5.1,2.1){11};
\node at (5.7,2.1){6};
\node at (6.3,2.1){2};
\node at (6.9,2.1){6};
\node at (7.5,2.1){3};
\node at (8.1,2.1){2};

\node at (3.3,0){0.1};
\node at (3.9,0){0.2};
\node at (4.5,0){0.3};
\node at (5.1,0){0.4};
\node at (5.7,0){0.5};
\node at (6.3,0){0.6};
\node at (6.9,0){0.7};
\node at (7.5,0){0.8};
\node at (8.1,0){0.9};

\draw (5.1,0.9) circle (0.3cm);
\draw (6.3,2.1) circle (0.3cm);
\draw (7.5,0.9) circle (0.3cm);
\draw (8.1,2.1) circle (0.3cm);

  \draw  (3,0.6)--(8.4,0.6);
   \draw  (3,1.2)--(8.4,1.2);
   \draw  (3,1.8)--(8.4,1.8);
    \draw  (3,2.4)--(8.4,2.4);

   \draw  (3,0.6)--(3,2.4);
   \draw  (3.6,0.6)--(3.6,2.4);
    \draw  (4.2,0.6)--(4.2,2.4);
     \draw  (4.8,0.6)--(4.8,2.4);
    \draw  (5.4,0.6)--(5.4,2.4);
    
     \draw  (6,0.6)--(6,2.4);
   \draw  (6.6,0.6)--(6.6,2.4);
    \draw  (7.2,0.6)--(7.2,2.4);
     \draw  (7.8,0.6)--(7.8,2.4);
    \draw  (8.4,0.6)--(8.4,2.4);

  \draw decorate [decoration={brace,amplitude=6pt},xshift=4pt,yshift=0pt] {(2.7,0.6) -- (2.7,2.4)};
  
  \node at (1.9,-1){$\lambda_{min}(\cdot) > \eta^i$};
  \node at (1.9,-1.7){$\lambda_{min}(\cdot) > 0.1$};
  \node at (1.9,-2.4){$\lambda_{min}(\cdot) > 0.4$};
  \node at (1.9,-3.1){$\lambda_{min}(\cdot) > 0.6$};
  \node at (1.9,-3.8){$\lambda_{min}(\cdot) > 0.8$};

    \node at (0.5,-1){$i$};
  \node at (0.5,-1.7){$1$};
  \node at (0.5,-2.4){$2$};
  \node at (0.5,-3.1){$3$};
  \node at (0.5,-3.8){$4$};

  \node at (6.5,-1){$(\lambda_{max}^i, \lambda_{min}^i)$};
  
  \node at (8.45,-1){$\kappa^i=\frac{\lambda_{max}^i}{\lambda_{min}^i}$};
   \node at (6.5,-1.7){$(1,0.4)$};
  \node at (6.5,-2.4){$(2,0.6)$};
  \node at (6.5,-3.1){$(2,0.8)$};
  \node at (6.5,-3.8){$(2,0.9)$};

  \node at (4.2,-1){min $\{\lambda_{max}(\cdot)\}$};
   \node at (4.3,-1.7){$1$};
  \node at (4.3,-2.4){$2$};
  \node at (4.3,-3.1){$2$};
  \node at (4.3,-3.8){$2$};

   \node at (8.3,-1.7){$2.5$};
  \node at (8.3,-2.4){$3.3$};
  \node at (8.3,-3.1){$2.5$};
  \node at (8.3,-3.8){$2.2$};

  \draw  (0.2,-1.4)--(9.1,-1.4);
  \draw  (0.2,-0.5)--(9.1,-0.5);
  \draw  (0.2,-0.55)--(9.1,-0.55);
  
  \draw  (0.2,-4)--(9.1,-4);
  \draw  (0.2,-4.05)--(9.1,-4.05);

\end{tikzpicture}
\label{Flowdia2}
\end{figure}

\vspace{-0.3cm}

\section{Acknowledgments}
This work was funded in part by TCS Research Scholarship Program under project number TCS/CS/2011191E and in part by SERB, Dept. of Science and Technology, GoI under project number SERB/EE/2017242.

\end{document}